%%%%%%%%%%%%%%%%%%%%%%%%%%%%%%%%%%%%%%%%%%%%%%%%%%%%%%%%%%%%%%%%
%%%%%%%%  Parafermionic exclusion principle.... 
%%%%%%%%  P. Jacob and P. Mathieu
%%%%%%%%%%%%%%%%%%%%%%%%%%%%%%%%%%%%%%%%%%%%%%%%%%%%%%%%%%%%%%%%

%\magnification1100

\input harvmac.tex

%\draft

%-------------------------------------------------------------------------------
% contractions de Wick
%
\def\ubrackfill#1{$\mathsurround=0pt
	\kern2.5pt\vrule depth#1\leaders\hrule\hfill\vrule depth#1\kern2.5pt$}
\def\contract#1{\mathop{\vbox{\ialign{##\crcr\noalign{\kern3pt}
	\ubrackfill{3pt}\crcr\noalign{\kern3pt\nointerlineskip}
	$\hfil\displaystyle{#1}\hfil$\crcr}}}\limits
}

\def\ubrack#1{$\mathsurround=0pt
	\vrule depth#1\leaders\hrule\hfill\vrule depth#1$}
\def\dbrack#1{$\mathsurround=0pt
	\vrule height#1\leaders\hrule\hfill\vrule height#1$}
\def\ucontract#1#2{\mathop{\vbox{\ialign{##\crcr\noalign{\kern 4pt}
	\ubrack{#2}\crcr\noalign{\kern 4pt\nointerlineskip}
	$\hskip #1\relax$\crcr}}}\limits
}
\def\dcontract#1#2{\mathop{\vbox{\ialign{##\crcr
	$\hskip #1\relax$\crcr\noalign{\kern0pt}
	\dbrack{#2}\crcr\noalign{\kern0pt\nointerlineskip}
	}}}\limits
}

\def\ucont#1#2#3{^{\kern-#3\ucontract{#1}{#2}\kern #3\kern-#1}}
\def\dcont#1#2#3{_{\kern-#3\dcontract{#1}{#2}\kern #3\kern-#1}}

%=================================================================
% MACROS

% TABLEAU NOIR (BLACKBOARD)
\font\tenmsy=msbm10
\font\sevenmsy=msbm10 at 7pt
\font\fivemsy=msbm10 at 5pt
\newfam\msyfam % family 11
\textfont\msyfam=\tenmsy
\scriptfont\msyfam=\sevenmsy
\scriptscriptfont\msyfam=\fivemsy
\def\blackB{\fam\msyfam\tenmsy}
\def\ZZ{{\blackB Z}}

\def\II{{\blackB I}}

\let\R\rangle

\let\vph\varphi
\let\da\dagger
\def\Dk{{\Delta_k}}
\def\Dkl{{\Delta_{k,\ell}}}
\def\ab{{\bar a}}

\def\frac#1#2{{\textstyle{#1\over #2}}}

% alignements multiples
\def\eqalignD#1{
\vcenter{\openup1\jot\halign{
\hfil$\displaystyle{##}$~&
$\displaystyle{##}$\hfil~&
$\displaystyle{##}$\hfil\cr
#1}}
}

\def\eqalignQ#1{
\vcenter{\openup1\jot\halign{
\hfil$\displaystyle{##}$~&
$\displaystyle{##}$\hfil~&
$\displaystyle{##}$\hfil~&
$\displaystyle{##}$\hfil~&
$\displaystyle{##}$\hfil\cr
#1}}
}

\def\text#1{\quad\hbox{#1}\quad}
\def\gh{\hat{g}}

\def\la{\lambda}

\def\a{\alpha}

\def\qb{\bar{q}}

\def\Ol{{\displaystyle\Om_\geq^\la}}
\def\Ola{{\displaystyle\Om_{\geq}^{\la_1}}}
\def\Olaa{{\displaystyle\Om_{\geq}^{\la_2}}}

\def\A{{\cal{A}}}
\def\B{{\cal{A}}^\dagger}

\def\y{{\infty}}

\let\Rw\Rightarrow

\def\gh{{\widehat g}}
\def\rw{\rightarrow}

\def\Om{\mathop{\Omega}\limits}

\def\R{\rangle}

\def\su{\widehat{su}}

%\def\E1{E_{1}}

% Equations (overrides harvmac's equation macros)
\newcount\eqnum
\eqnum=0
\def\eq{\eqno(\secsym\the\meqno)\global\advance\meqno by1}
\def\eqlabel#1{{\xdef#1{\secsym\the\meqno}}\eq }

% References (overrides harvmac's reference macros)
\newwrite\refs
\def\startreferences{
 \immediate\openout\refs=references
 \immediate\write\refs{\baselineskip=14pt \parindent=16pt \parskip=2pt}
}
\startreferences

\refno=0
\def\aref#1{\global\advance\refno by1
 \immediate\write\refs{\noexpand\item{\the\refno.}#1\hfil\par}}
\def\ref#1{\aref{#1}\the\refno}
\def\refname#1{\xdef#1{\the\refno}}

\newcount\exno
\exno=0
\def\Ex{\global\advance\exno by1{\noindent\sl Example \the\exno:

\nobreak\par\nobreak}}

\parskip=6pt

\overfullrule=0mm

%%%%%%%%%%%%%%%%%%%%%%%%%%%%
\def\frac#1#2{{#1 \over #2}}

\def\suh{{\widehat {su}}}
\def\uh{{\widehat u}}
\def\rw{{\rightarrow}}

% References (overrides harvmac's reference macros)
\newwrite\refs
\def\startreferences{
 \immediate\openout\refs=references
 \immediate\write\refs{\baselineskip=14pt \parindent=16pt \parskip=2pt}
}
\startreferences

\refno=0
\def\aref#1{\global\advance\refno by1
 \immediate\write\refs{\noexpand\item{\the\refno.}#1\hfil\par}}
\def\ref#1{\aref{#1}\the\refno}
\def\refname#1{\xdef#1{\the\refno}}

%===============================================================================

% PAGE TITRE
\Title{\vbox{\baselineskip12pt
\hbox{
%LAVAL-PHY-{00-21}
}}}
{\vbox {\centerline{Parafermionic quasi-particle basis and fermionic-type characters }}}

\smallskip
\centerline{ P. Jacob and P. Mathieu
% \foot{Work supported by NSERC (Canada) and FCAR (Qu\'ebec) }
% \foot{Corresponding author (PM): tel:
% 418-656-3416; fax: 418-656-2040 }
}

\smallskip\centerline{ \it D\'epartement de
physique,} \smallskip\centerline{Universit\'e Laval,}
\smallskip\centerline{ Qu\'ebec, Canada G1K 7P4}
\smallskip\centerline{(pjacob@phy.ulaval.ca, pmathieu@phy.ulaval.ca)}
\vskip .2in
\bigskip
%\bigskip
%\bigskip
%\bigskip
\bigskip
\centerline{\bf Abstract}
\bigskip
\noindent
 
A new basis of states for highest-weight  modules in $\ZZ_k$ parafermionic conformal theories is displayed. It
is formulated in terms of an effective exclusion principle constraining strings of
$k$ fundamental parafermionic modes. The states of a module are then built by a simple
filling process, with no singular-vector subtractions. That results in fermionic-sum representations of the
characters, which are exactly the Lepowsky-Primc expressions.  We also stress that the
underlying combinatorics
-- which is the one pertaining to the Andrews-Gordon identities -- has a remarkably natural parafermionic
interpretation.

%Classification numbers: 11.10.-z, 02.20+bm

%Keywords: conformal field theory, parafermions, 
%$Z_N$ models; character; singular vectors.
\Date{06/01\ }
%\ (hepth@xxx/0108063)}

\let\n\noindent

%==============================================================================

\newsec{Introduction}

\subsec{Bosonic- vs fermionic-sum representation of characters}

Over the last years, there has been a great amount of activities centered on novel -- i.e., fermionic-like -- 
representations of the characters in rational conformal field theories.  The standard expression for
the irreducible characters is obtained by the subtraction of the singular vectors in reducible
Verma modules  (which is usually an infinite sequence of subtractions and additions).  The
prototype of such character expressions are the well-known Rocha-Caridi
formulae   for irreducible highest-weight modules
$|h_{rs}\R$, 
$h_{rs}= [(pr-p's)^2-(p-p')^2]/4pp'$,  in Virasoro minimal models [\ref{A. Rocha-Caridi, in {\it Vertex
Operators in Mathematics and Physics}, Eds  J. Lepowsky et al, Publ. Math. Sciences Res. Inst. 3,
Springer-Verlag, (1985) 451.}]:
$$ \chi_{rs}(q)= { q^{-{1/24}} \over (q)_\y} \sum_{n \in \ZZ} \left[q^{(2pp'n+ pr- p's)^2/4pp'}- q^{(2pp'n+
pr+ p's)^2/4pp'}\right]
\eqlabel\rocha$$ 
We used the following notation  for  the  Euler function:
$$(q)_\y = \lim_{n\rw\y} (q)_n\; , \qquad   (q)_n= \prod_{i=1}^n (1-q^i) \eqlabel\dedeqn$$
with $(q)_0=1$.  
The Kac-Peterson characters of the integrable representations in affine Lie
algebras provide the analogue for WZW models [\ref{V. Kac and D. Peterson, Adv. Math. {\bf 53} (1984)
125.}\refname\KP]. 

 Characters of that type are termed `bosonic' in [\ref{R. Kedem, T.R.
Klassen, B. M. McCoy and E. Melzer, Phys. Lett. {\bf B304} (1993) 263.}\refname\KKMM] (see also [\ref{E.
Melzer, Int. J. Mod. Phys. {\bf A9} (1994) 133.}\refname\Mel]) since they
can be obtained by a BRST projection of free-field (typically, bosonic) representations (cf. [\ref{G. Felder,
Nucl. Phys. {\bf B317} (1989) 215}] for the minimal models). More precisely, such characters are  obtained by
the free application of the various lowering operators of the underlying  infinite algebra and the subsequent
corrections resulting from the subtraction of the singular vectors. For the Virasoro case, the ladder
operators are the
$L_{n<0}$ modes and their free action is signaled in (\rocha) by the presence of the factor
$(q)_\y^{-1}$. For affine Lie algebras, these are the
 operators that implement  the subtraction of positive roots.  

In brief, a bosonic character reflects the
construction of an irreducible  module starting from a highest-weight Verma module (by which we mean a module
generated freely by the application of the lowering operators). Its hallmark are thus not only the
$(q)_\y^{-1}$ factors but also the alternating signs in the character expression. 

By contrast, `fermionic-type' characters or more precisely, fermionic-sum representations have a quite
different structure. Unsurprisingly, the prototype is the character of a free-fermion Fock space whose natural
product form has the following sum decomposition:
$$
 \chi^{{\rm ff}} ={q^{-1/48}}\prod_{n=0}^\y (1+q^{n+1/2}) = \sum_{n=0} {q^{n^2/2}\over
(q)_n}\eqlabel\isingca$$ The rhs  is the generating function for partitions in
distinct parts -- which is appropriate for fermions -- and this sum representation is obtained by a simple
variant of the Euler identity (cf. eq. (2.2.6) of corollary 2.2 of [\ref{G.E. Andrews, {\it The theory of
partitions}, Cambridge Univ. Press (1984).}\refname\Andr]):
$$ \prod_{n=0}^\y (1+tq^{n})= \sum_{n=0}^\y {t^nq^{n(n-1)/2}\over (q)_n}\eqlabel\euler$$ with $t=q^{1/2}$. The
sum
 describes the filling of the space of states in terms
of  particles -- each term of the sum representing the contribution
of a different particle-number sector --, subject to a restriction rule, here the Pauli exclusion principle.
This interpretation of the filling of the space of states in terms of quasi-particles with restriction rules is
the characteristic feature of fermionic-sum representations.  It generically manifests itself by the presence
of
$(q)_n$ factors in the denominator. But more significantly, it is expressed in the form of a positive $q$
series (i.e., there are no alternating signs).  The bosonic counterpart of this fermionic sum  is simply the
combination of the  Rocha-Caridi characters
$\chi_{11}+\chi_{21}$. 

As another simple example, consider the free-boson theory, whose character is  $q^{-1/24}(q)_\y^{-1}$,
and its representation in terms of two complex fermions. Each complex fermion is described by the partition
${q^{-1/48}}\prod_{n=0}^\y (1+q^{n+1/2})$. However, in order to keep track of the charge, we introduce an
auxiliary label $z$ and represents the two characters as ${q^{-1/48}}\prod_{n=1}^\y (1+zq^{n-1/2})$ and
${q^{-1/48}}\prod_{n=0}^\y (1+z^{-1}q^{n+1/2})$ instead.  Our free boson theory is recovered from the
zero-charge sector of the product of these characters. Using again the Euler identity (\euler),
we recover, from this projected product, another well-known  identity of Euler (cf. eq. (2.2.5) of corollary
2.2 of [\Andr]): 
$${1\over (q)_\y}= \sum_{n=0}^\y {q^{n^2}\over (q)_n^2}\eqlabel\bosca$$
The rhs is the fermionic-sum representation of a free-boson character (up to the $q^{-1/24}$ factor). For this
example, the bosonic counterpart, namely the lhs, is rather simple in that there are no singular-vector
subtractions. In other words, the bosonic-sum representation reduces to a single term.  But the `bosonic
identification mark'
$(q)_\y^{-1}$ is present.

Let us consider still another example, namely the Virasoro characters of the Yang-Lee model $(p',p)=(2,5)$,
which reads
$$\chi_{1,1+a}=  q^{(11-12a)/60}\sum_{m=0}^\y {q^{m(m+a)}\over (q)_m} =
q^{(11-12a)/60}\prod_{n=1\atop n\not=0,(2-a)\,{\rm  mod}  \, 5}^\y{1\over (1-q^n)}
% (1-q^{5n-4+a})(1-q^{5n-1-a})}
\eqlabel\yang$$ with
$a=0,1$. The above equalities are nothing but the famous Rogers-Ramanujan-Schur identities (cf. [\Andr],
eqs (7.1.6), (7.1.7)).\foot{The Lie algebraic
derivation of this relation is given in Ex. 14.8 of [\ref{P. Di Francesco, P. Mathieu and D. S\'en\'echal, {\it
Conformal Field theory}, Springer Verlag, 1997.}\refname\DMS] and appears to be due to Kac [\ref{V. Kac,
{\it Modular invariance in mathematics and  physics}, address at the centennial of the AMS,
(1988).}\refname\Kac]. The product form of these Virasoro characters was discovered independently in
[\ref{P. Christie, Int. J. Mod. Phys. {\bf A6} (1991) 5271.}].}    As indicated, this fermionic-sum
representation has a natural product version. This result has a natural  generalization to the whole
$(p',p)=(2,2k+1)$ series.
\foot{Are there  analogous bosonic vs fermionic forms for the character expressions in the case of finite
algebras? Although a bit far-fetched, the following
$su(2)$ Lie-algebra example indicates that this is indeed so, at least if `fermionic representation' is
understood in the broader sense of `quasi-particle representation'.  For an irreducible representation with
finite Dynkin label
$\ell$ (i.e., $\ell= 2j$), which has dimension $\ell+1$, the Weyl character formula yields
$$\chi_{(\ell)} = { q^{-\ell}-q^{\ell+2} \over 1-q^{2} }= {q^{-\ell}(1-q^{2\ell+2})\over (q^{2})_1 } $$
This is the bosonic form of the character, that captures the subtraction of one singular vector. On the other
hand, the Schwinger's oscillator model of angular momentum algebra, that is, the representation of the algebra
in terms of two independent harmonic oscillators (see e.g., [\ref{J.J. Sakurai, {\it Modern quantum
mechanics}, Addison Wesley (1994) sect. 3.8.}]),  yields
the quasi-particle representation:
 $$\chi_{(\ell)}= q^{-\ell} \sum_{m=0\atop m-\ell=0 \,{\rm mod} 2}^\ell q^{2m}= q^{-\ell} \sum_{m=0\atop
m-\ell=0 \,{\rm mod} 2}^\ell {(q^2)^m\over (q^2)_0}$$
This is obtained by projecting the product of the charged harmonic oscillator characters $(1-zq)^{-1}
\,(1-z^{-1}q)^{-1}$ onto the $z^{2\ell}$ sector. In this case however, the quasi-particles have no fermionic
attribute.
}

\subsec{The origins of fermionic-sum  characters in conformal field theory}

The discovery, over the last decade,  of a rich class of fermionic-sum representations in conformal field 
theory  has various origins.

\n {\it A- Representation theory of Lie algebras}

The first fermionic formulae for conformal characters appeared in
mathematics, as the characters of the $Z$  Lepowsky-Wilson algebra [\ref{J. Lepowsky and R.L. Wilson, Inv.
Math. {\bf 77} (1984) 199.}], which turns out to be equivalent to the parafermionic algebra of
Zamolodchikov-Fateev [\ref{A.B.
Zamolodchikov and V.A. Fateev, Sov. Phys. JETP {\bf 43} (1985) 215.}\refname\ZFa]. The resulting characters
have been obtained in [\ref{J. Lepowsky and M. Primc,  Contemporary
Mathematics {\bf 46} AMS, Providence, 1985.}\refname\LP]. By construction, these characters are associated
to those of the primary fields of  the coset model $\suh(2)_k/\uh(1)$, for which bosonic-type formulae already
existed - these are the string functions of [\KP].  That provided the first nontrivial example of
bosonic-fermionic character identities.  
Both sides (bosonic vs fermionic) have direct $\suh(N)$ extensions and these are related to generalized
parafermionic models [\ref{D. Gepner, Nucl. Phys. {\bf B290} (1987) 10.}].

The Lepowsky-Primc formulae, in a slightly modified form, pop up in a second and quite unexpected
conformal-field-theoretical context, namely as  the characters of
$(2,2k+1)$ minimal models [\Kac, \ref{B.L. Feigin, T. Nakanishi  and H. Ooguri, Int. J. Mod. Phys. {\bf A7}
Suppl. {\bf 1A} (1992) 217.}\refname\FNO].  For $k=2$, this is related to the result (\yang).

\n {\it B- Dilogarithm identities}

A different development  
is rooted in the dilogarithm identities,  originating from the TBA
analysis of integrable perturbations of conformal field theory [\ref{T.R. Klassen and E. Melzer, Nucl. Phys.
{\bf B338} (1990) 485; {\bf B370}
(1992) 511.}],  relating the central charge and the
conformal dimensions.\foot{Such identities also arise from the low temperature limit of the free energy of
integrable lattice systems (RSOS-type spin chains) [\ref{V.V. Bazhanov and N. Yu Reshetikhin, Int. J. Mod.
Phys. {\bf A4} (1989) 115; A. N. Kirillov, J. Sov. Math. {\bf 47} (1989) 2450; A. N. Kirillov and N. Yu
Reshetikhin, J. Sov. Math. {\bf 52} (1989) 3156.}] and in the leading behavior of the lattice transfer matrix
[\ref{A. Kl\"umper and P. A Pearce, J. Stat. Phys. {\bf 64} (1991) 13.} ]. For a beautiful review of the
various facets of the dilogarithm identities, see [\ref{A.N. Kirillov, 
{\it Dilogarithm identities},  Prog. Theo  Phys. Suppl. 118 (1995) 61-142.}].} Within the scope of  conformal
field theory, these identities can be obtained by comparing the asymptotic behavior of the usual bosonic
character expressions with their fermionic-sum representations [\ref{W. Nahm, A. Recknagel and M. Terhoeven,
Mod. Phys. Lett. {\bf A8} (1993) 1835.}\refname\NRT], which were known for the $(2,2k+1)$ models.
It has then been suggested that such identities (which are not limited to the $(2,2k+1)$ series) could  somehow
be lifted to the whole characters [\ref{M. Terhoeven, {\it Lift of dilogarithm to partition identities},
hep-th/9211120.}\refname\Ther], producing
 characters written naturally as fermionic sums, as in the rhs of the Andrews-Gordon identity
[\ref{G.E. Andrews, Proc. Nat Acad. Sci. USA {\bf 71} (1974) 4082.}\refname\And, \ref{B. Gordon, Amer. J. Math.
{\bf 83} (1961) 393.}\refname\Gor] 
$$\prod_{n\not=0,\pm i\,{\rm mod}\, 2k+1} {1\over 1-q^n}= \sum_{m_1,\cdots , m_{k-1}}{\ q^{N_1^2+\cdots
 N_{k-1}^2+N_i+\cdots +N_{k-1}}\over (q)_{m_1}\cdots (q)_{m_{k-1}} }\eqlabel\gordon$$

This program has effectively been realized in [\ref{A. Kuniba, N. Nakanishi and J. Suzuki, Mod. Phys. Lett.
{\bf A 8} (1993) 1649.}]. Building on an earlier observation [\ref{A. Kuniba and  N. Nakanashi,
Mod. Phys. Lett. {\bf A 7} (1992) 3487.}] that the complete spectrum of generalized $\gh/\uh(1)^r$
parafermionic conformal theories steams from the dilogarithm function, 
the authors have  conjectured a natural lift of
the asymptotic parafermionic characters in the form of fermion-sum representations that include [\LP] as a
special case.

\n {\it C- The scaling limit of spin chains}

At the same period, fermionic-sum representations emerge from a completely different perspective. In
[\ref{R. Kedem and B. McCoy, J. Stat. Phys {\bf 71} (1993) 883.}, \ref{S. Dasmahapatra, R.
Kedem, B. M. McCoy and E. Melzer, J. Stat. Phys {\bf 74} (1994) 239.}], the spectrum of the gapless
three-state Potts model has been computed in the conformal limit by solving the Bethe equations. That leads
to a description of the hamiltonian eigenspectrum in terms of quasi-particles obeying a fermionic exclusion
principle in momentum space, supplemented by further restrictions constraining the momentum eigenvalues. These
spectrum computations lead to characters expressed directly as fermionic sums. In the (anti)
ferromagnetic sector, the scaling limit of this model is the ($\ZZ_4$)
$\ZZ_3$ parafermionic theory and the built characters agree with the results of [\LP].  A key
observation at this stage appeared to be the rewriting of the Lepowsky-Primc result in terms of the inverse
Cartan matrix for
$\su(k)$, which, in turn, suggested a large number of natural generalizations [\KKMM, \ref{R. Kedem, T.R.
Klassen, B. M. McCoy and E. Melzer, Phys. Lett. {\bf B307} (1993) 68.}\refname\KKMMa, \ref{S. Dasmahapatra, R.
Kedem, T.R. Klassen, B. M. McCoy and E. Melzer, Int. J. Mod. Phys. {\bf B7} (1993) 3617.}\refname\DKKMM, 
\ref{R. Kedem, B. M. McCoy and E. Melzer, in {\it Recent progress in statistical mechanics and quantum field
theory}, ed. by P. Bouwknegt et al, World Scientific, (1995) 195. }\refname\KMM], all having an underlying
quasi-particle interpretation.

The proof of most these identities has been supplied in the following years. An important step in that
direction was the observation [\Mel] that the Virasoro characters have a natural `finitized version' in terms
of path spaces -- or corner-transfer-matrix sums -- in the RSOS model [\ref{G.E. Andrews, R.J. Baxter and P.J.
Forrester, J. Stat. Phys. {\bf 35} (1984) 193.}].  That allowed for a proof of the simplest cases;  this
line of attack  has subsequently been developed vigorously, starting with the paper [\ref{A. Berkovich, Nucl.
Phys. {\bf B415} (1994) 691.}].

%** le reste de cette sous section est nouveau

Few years later, the fermionic-type exclusion principle underlying the quasi-particle filling of the space of
states was shown in [\ref{A. Berkovich and B.M. McCoy, 
%{\it The universal chiral partition function for
% exclusion statistics}, hep-th/9808013.}
in {\it Statistical physics at the eve of the 21st century}, 
Series on Adv. in Stat. Mech., vol 14, ed. by M.T Batchelor and L.T. White, World Scientific, 1999.}] to be precisely the Haldane's generalization of the Pauli exclusion
principle [\ref{F.D.M. Haldane,  Phys. Rev. Lett. {\bf 67} (1991)  937.}\refname\hal]. 

\n {\it D- Spinons bases and generalization}

Quasi-particle descriptions of conformal field  theories also originate from still another and quite different
source. This one is rooted in the discovery of the remarkable connection between the $su(N)$ 
Haldane-Shastry model [\ref{F.D.M. Haldane, Phys. Rev. Lett. {\bf 60} (1988) 635; B.S.  Shastry, Phys. Rev.
Lett. {\bf 60}  (1988) 639.}] (an integrable spin chain with long-range interaction) and $\su(N)_1$ WZW
model, the former appearing as 
 a sort of  discretization of the latter [\ref{F.D.M. Haldane, Z.N.C. Ha, J.C. Talstra, D. Bernard and  V.
Pasquier, Phys. Rev. Lett. {\bf 69} (1992)  2021.}\refname\pasquier]. This relation  yields naturally a 
spinon description of the WZW basis of states, which has been
lifted to fermionic-sum representations for the characters  [\ref{D. Bernard, 
V. Pasquier and
D. Serban, Nucl. Phys. {\bf B 428} (1994) 612.}\refname\serban, \ref{ P. Bouwknegt. A.A. Ludwig and K. 
Schoutens, Phys. Lett. {\bf B 338} (1994) 448; 
{\bf B 359} (1995) 304. }\refname\schou]. 

As an outgrowth of the spinon descriptions of WZW models, a  general approach for quasi-particle
reformulation of conformal field theories has been presented in [\ref{K.  Schoutens, Phys. Rev. Lett. {\bf 79}
(1997) 2608.}] (and much exemplified in e.g., in [\ref{ P. Bouwknegt. and K. 
Schoutens, Nucl. Phys. 
{\bf B 547} (1999) 501. }\refname\bouschou]). It  revolves around the construction of truncated
fermionic-type characters.

\subsec{Bosonic vs fermionic  parafermionic characters}

Let us return to the parafermionic case. Recently, we have obtained new expressions for the $\ZZ_k$
parafermionic characters [\ref{P. Jacob and P. Mathieu, Nucl. Phys. {\bf B 587} (2000) 514.}\refname\JM].
These  expressions are different from those obtained by the coset construction, the string
functions,  themselves identical to the parafermionic  characters obtained by the BRST projection of a
two-boson representation  [\ref{J. Distler and Z.  Qiu,  Nucl.Phys. {\bf B336} (1990) 533.}\refname\DQ,
\ref{D. Nemeschansky, Nucl. Phys. {\bf B363}  (1989) 665.}\refname\Nem].  These new character formulae are
rooted in the description of the parafermionic  states in terms of a standard basis (cf. section 2). The
highest-weight modules constructed in the framework of this  basis contain an
infinite number of singular vectors. The explicit expression of all these singular vectors has been obtained
and by their appropriate subtraction and addition, we have constructed 
 bosonic-type  parafermionic characters [\JM].\foot{Actually, they do not have a purely bosonic
appearance in that they have no $(q)_\y^{-1}$ factors and they contain fermionic ingredients, namely $(q)_n$
terms in the denominators -- cf. appendix A. However, their alternating-sign structure qualifies them as
genuine bosonic sums. Note that their direct analytic equivalence with the string functions has not been
established yet.}  
The resulting formula is reported
in appendix A in a somewhat improved form. 

Given these new expressions, it is natural to try to establish directly their equivalence to
the ones in [\LP]. However, a frontal  attack is difficult, even in the two extreme cases $k=2$
(the Ising model) and the
$k=\y$ (the two complex bosons limit - cf. appendix B). That suggested us to look instead for a reformulation
of our standard basis in the form of a new fermionic-type basis that would be expressed in terms of the modes
of a single type of parafermionic field complemented with specific restriction rules. The aim of this paper is
to present that new basis, derived from a purely parafermionic perspective.

We complete this introduction with a brief presentation of the paper's content.
The next section is intended to fix the notation and present the standard basis of [\JM].  
The new basis is displayed in section 3. It is formulated in terms of simple restriction
rules.  These rules are first motivated by inductive considerations  and heuristic arguments relying on an
exclusion principle for
$\ZZ_k$ quasi-particles. The
selection rules are then shown to be rooted in the $\ZZ_k$ invariance. Technical details
establishing the genuine character of the proposed basis are reported in an appendix.  The basis  is then
illustrated for various simple modules in section 4. In section 5, this basis is lifted to the general
expression of the
$\ZZ_k$ characters, written in terms of a positive sum over restricted partitions. This result is
then reexpressed in product form using Andrews identity. We obtain complete agreement with the Lepowsky-Primc
formula [\LP].

%------------------------------------------------------
\newsec{The standard basis of the $\ZZ_k$ parafermionic modules}

The  $\ZZ_k$ parafermionic conformal algebra [\ZFa, \ref{A.B.
Zamolodchikov and V.A. Fateev, Sov. Phys. JETP  {\bf 63} (1986) 913.}\refname\ZFb]
is generated by the parafermionic fields $\psi_r= \psi^\dagger_{k-r}$, $r=0,1,\cdots ,
k-1$, of conformal
dimension 
$  h_{\psi_r}={r(k-r)/ k}$. Since the higher-order parafermions and their conjugate can be
obtained from composition of the fundamental ones, $\psi_1$ and $\psi_1^\da $, it suffices to
write the commutation relations involving the modes of these fundamental parafermions. 
The mode decomposition of $\psi_1$ and $\psi_1^\da $ acting on a generic field of charge
$\phi_q$ reads:
$$\eqalign{
 \psi_1(z)\phi_{q}(0) & = \sum_{m=-\y}^\y
z^{-q/k-m-1}A_{(1+q)/k+m}\, \phi_{q}(0)\cr
\psi_1^\dagger (z)\phi_{q}(0)  &= \sum_{m=-\y}^\y
z^{q/k-m-1} A^\dagger_{(1-q)/k+m}\,\phi_{q}(0)\cr}
\eqlabel\modep$$
the fractional power of $z$ being fixed by the mutual locality [\ZFa].
Note that the
dimension of a parafermionic mode is the negative of its index.

Since the fractional
part of the modes is fixed unambiguously by the charge of the field or 
state on which it acts, it can be omitted.  
To indicate that the fractional part has been deleted (which implies that 
the conformal dimension of the mode is no longer given by minus its
index), we use calligraphic symbols,\foot{This is the notation used
in [\JM] except
that there
${\cal B}$ was used to denoted $\A^\dagger$ and the symbol $N$ played the role of $k$.} 
i.e., 
$$\A_n |\phi_{q}\R\equiv  A_{n+(1+q)/k}|\phi_{q}\R\, ,\qquad
\A_n^\dagger|\phi_{q}\R\equiv 
A_{n+(1-q)/k}^\dagger|\phi_{q}\R\eq$$
Given that the parafermions $\psi_1, \,\psi_1^\da $ have respective charge 2 and $-2$, the modes
$\A_n $ and $
\A_n^\dagger$ also have charge $2$ and $-2$. When reinserting the fractional part of the
modes in a given string of operators, these charges must be taken into account. For instance, we have
$$(\A_{-1})^3 |\phi_{q}\R\equiv  A_{-1+(5+q)/k}A_{-1+(3+q)/k}A_{-1+(1+q)/k}|\phi_{q}\R\eq$$

With this simplifying notation, the commutation relation reads then [\ZFa, \JM]
$$ \eqalign{ &\sum_{l=0}^{\infty} C^{(l)}_{-2/k-1} \left[ \A_{n-l-1}
\B_{m+l+1} + \B_{m-l}\A_{n+l} \right] \phi_{[q,\qb]}
(0) \cr  &=\left[ {(k+2) \over k} L_{n+m} + {1 \over 2}
(n+{q\over k})(n-1+{q\over k})   
\delta_{n+m,0} \right] \phi_{[q,\qb]}(0)\cr} 
\eqlabel\comasi$$ 
and
$$ 
 \sum_{l=0}^{\infty} C_{2/k}^{(l)} 
\left[ \A_{n-l}\A_{m+l}-\A_{m-l}\A_{n+l} \right]
\phi_{[q,\qb]}(0) =0 \eqlabel\combsi$$
together with a similar relation with $\A\rw \A^\da$ and 
where $ C_t^{(l)}=  {\Gamma(l-t)/ l!\,
\Gamma(-t)}$.

In order to describe the standard basis of states, we need to recall the definition
of a parafermionic highest-weight state [\ZFa]. For a fixed value of $k$, there are $k$
parafermionic primary fields 
$\{\vph_\ell\,| \ell= 0,\cdots ,k-1\} $;  $\vph_\ell$ has charge $\ell$ and
dimension
$  h_{\ell}={\ell(k-\ell) / 2k(k+2)}$.
 To each primary field, there
corresponds a highest-weight state
$|\vph_\ell\R$, with $|0\R=|\vph_0\R$.  The parafermionic highest-weight  
conditions are
$$\A_{n} | \vph_\ell 
\rangle = \A^\dagger_{n+1} | \vph_\ell \rangle 
=0  \qquad {\rm for}\quad n\geq 0\eqlabel\hiwe$$

The set of independent states at level $s$ in the
highest-weight module of
$|\vph_\ell\R$ of charge $\ell+2r$ is generated by the following states [\JM]:
$$\A_{-n_1}\A_{-n_2} .. .  \A_{-n_{j}}\B_{-m_1}\B_{-m_2} .. .  \B_{-m_{p}}
|\vph_\ell\R\eqlabel\mixs$$
with
$$ j-p=r\, , \quad n_{i} \geq n_{i+1}\geq 1\,, \qquad m_{i} \geq m_{i+1}\geq 0\,,
\qquad
\sum_{i=1}^{j} n_i+
\sum_{i=1}^{p} m_i=s\eqlabel\mixst$$
In the standard basis, we can thus order separately the $\A$ strings and the
$\B$ strings but there is no mixed ordering. 

The highest-weight module of $|\vph_\ell\R$ can be decomposed into a direct sum of $k$ modules of fixed charge
defined modulo
$2k$.  In the
highest-weight module of charge $\ell+2r$ (or relative charge $2r$), the highest-weight state  is
$$ |\varphi_\ell^{(r)}\R = (\A_{-1})^r| \varphi_\ell\R  \qquad (0\leq r\leq
k-\ell)\eqlabel\bb$$ 
Similarly, if $r<0$, the charged highest-weight state reads
$$|\varphi_\ell^{(-r)}\R = (\B_{0})^r|\varphi_\ell\R \qquad 0\leq r \leq
\ell\eqlabel\bc$$ 
The upper bounds for the values of $r$ are induced by the presence of the `primary' singular vectors:
$$|\chi_\ell\R\ = \A_{-1}^{k-\ell+1}|\vph_\ell\R \, ,\qquad
|\chi'_\ell\R\ = (\B_{0})^{\ell+1}|\vph_\ell\R \eqlabel\sig$$

 Not all the states $|
\varphi_\ell^{(\pm r)}\R$ are independent however; there is the following field
identification [\ref{D. Gepner and Z. Qiu, Nucl. Phys. {\bf B285} [FS19] (1987)
423.}\refname\GQ]:
$$\vph_{\ell}^{(-\ell)}\sim \varphi^{(k-\ell)}_\ell \sim
\vph_{k-\ell}\equiv \varphi_\ell^\dagger \eqlabel\fide$$
This implies that either
$\{\vph_{\ell}^{(r)}\}$ or
$\{\vph_{\ell}^{(-r)}\}$  can be chosen as a complete set of basic parafermionic fields (which  correspond to
the highest weights in a fixed charge module). In other words, a complete set of charged  highest-weight
states is generated from the parafermionic primary fields by applying either strings of $\A$ or $\A^\da$
operators; at this level, the use of both types of modes is
thus superfluous. As it will be shown in the next section, this can also be made true for all the descendants.

\newsec{A new parafermionic basis and the underlying exclusion principle}

\subsec{The quasi-particle basis}

We now look for a fermionic-type  basis for the parafermionic theories. By a
fermionic-type  basis, we mean a basis that is free of singular vectors, akin to the free-fermion basis, 
and in which
the states in a module are filled subject to simple restrictions. More
precisely, singular vectors should be eliminated by a selection rule rather than being subtracted.

Since  there is an infinite number of singular vectors in the standard basis, it would appear highly
improbable that they can all be taken into account by means of simple restrictions. This suggest that a
fermionic-type basis will not be naturally formulated in terms of both the $\A$ and $\A^\da$ modes but
rather in terms of a single type, say $\A$. This is certainly possible given that  
 $\psi_1^\da= \psi_{k-1}\sim ({\psi_1})^{k-1}$. In a single mode-type basis, the constraint of a
fixed value for the difference on the number of
$\A$ and
$\A^\da$ operators in a module of fixed charge is replaced by the conservation of the number of $\A$ operators
modulo
$k$.
%  (which reflects the  built-in $\ZZ_k$ invariance 
% thats translates into the condition
% $(\psi_1)^k\sim\II$).  

In a fermionic-type basis expressed in terms of the $\A$ modes only, states in the highest-weight module of 
$|\varphi_\ell\R$ of relative charge
$2r$ will naturally be written as
ordered strings of the form
$$\A_{-n_1}\A_{-n_2}\cdots \A_{-n_m}|\varphi_\ell\R\qquad \quad {\rm with}\quad 
n_i\geq n_{i+1}\geq 1\eqlabel\ferbas$$  
with $m=r$ mod $k$. In addition, we have to take into account the presence of the singular vector
$|\chi_\ell\R$ -- cf. (\sig).  This gives an upper bound on the possible number of $\A_{-1}$ factors at the
right end of the string.  Since the $n_i$'s are ordered and all greater or equal to 1,  this requirement is
fully captured by the simple condition:
$$n_{m-k+\ell}\geq 2\eqlabel\ferbass$$
That ensures that the $k-\ell+1$-th operator counted from the right is not $\A_{-1}$.
The set (\ferbas) with the restriction (\ferbass) still generates too much states, meaning that
the  specification of the basis requires further selection rules.

The missing selection rules are the following. All ordered sequences of $\A$ operators that contain any one of
the following $k$-string
$$(\A_{-(n+1)})^{k-i}(\A_{-n})^{i}\qquad (i=0,\cdots , k-1)\eqlabel\parexcl$$
must be forbidden. 
Together with the condition 
(\ferbass), these restrictions imposed on an ordered sequence of operators provide a basis.  
 Note that
when $\ell=0,1$, the condition (\ferbass) is superfluous, being  taken into account by the above restrictions.

In view of writing down the characters, that is, of counting the number of states per level, it is
convenient to reformulate the above restriction rules algebraically as follows:
$$   n_j- n_{j+k-1}\geq 2\eqlabel\restru$$
It is somewhat surprising that this simple condition  captures precisely the whole set of restrictions encoded
in (\parexcl).

The next two sections serve to establish the naturalness of the restrictions rules (\parexcl), using first
an inductive approach and then a sort of generalized exclusion principle.  The restriction rules are then
traced back to the mere $\ZZ_k$ invariance. That precisely the states (\parexcl) can be eliminated as a
result of the $\ZZ_k$ invariance is proved in appendix C.

\subsec{Inductive derivation of the restriction rules}

To justify the naturalness of the restriction rules (\parexcl), we first  `derive' them following an
inductive line.  The $k=2$ parafermionic theory is a free-fermion model. A natural starting point is thus to
first retranslate the fermionic Fock basis into the $\A$-mode language, and then look for the generalization
to 
$k>2$. 

Since the parafermionic field  is a free fermion when $k=2$, the standard fermionic
modes
$b_{-\nu}$ are nothing but the
$A_{-\nu}$ modes (keeping track of their fractional part - cf. [\JM] appendix C), e.g.,
$$b_{-n+1/2}|0\R =A_{-n+1/2}|0\R = \A_{-n}|0\R\eq$$
The fermionic Fock  basis is
$$b_{-\nu_1}b_{-\nu_2}\cdots b_{-\nu_m}|\varphi_\ell\R\qquad \quad {\rm with}\quad 
\nu_i> \nu_{i+1}\eqlabel\ferm$$
where $\ell=0,1$.
With 
$\nu_i=n_i-f_i$,
where $f_i$ stands for the fractional part,
it is directly translated into
$$A_{-n_1+(1+\ell+2p-2)/2}\cdots A_{-n_m+(1+\ell)/2}|\varphi_\ell\R\qquad{\rm
with}\quad  n_i-f_i> n_{i+1}-f_{i+1}\eqlabel\fermm$$
After being reexpressed in terms of the integer-mode operators $\A$, in the form (\ferbas), we see
that the ordering will be preserved if
$n_i>n_{i+1}$.  The fermionic nature of the $\A$ modes is itself taken into account by the condition
$n_i>n_{i+1}+1$.  
In other words, in any ordered sequence of $\A$ modes, we exclude the following
two subsequences:
$$\A_{-(n+1)}\A_{-n} \qquad {\rm and} \quad \A_{-n}\A_{-n} \eqlabel\exdeu$$
 The $k=2$ version of
our new basis will  then be spanned by all states of the form
$$\A_{-n_1}\A_{-n_2}\cdots \A_{-n_m}|\varphi_\ell\R\qquad \quad {\rm with}\quad 
n_i\geq n_{i+1}+2\eqlabel\fermicas$$
Note that the condition (\ferbass) is taken into account by the above difference 2 condition.

Consider now the $k=3$ model. The crucial step is finding the proper specification of the 
sequences that should be forbidden.  For $k=3$, these will be sequences of three $\A$ modes. The most
natural generalization of the exclusion (\exdeu) amounts to forbid
 the following 3-strings
$$\A_{-n}\A_{-n}\A_{-n}\;\;,\qquad \A_{-(n+1)}\A_{-n}\A_{-n} \;,\qquad
\A_{-(n+1)}\A_{-(n+1)}\A_{-n}\eqlabel\extri$$
These are obtained by inserting $\A_{-n}$ or $\A_{-(n+1)}$ at either end of (\exdeu).
Similarly, the appropriate $k=4$ forbidden $4$-states are:
$$\eqalignD{
& \A_{-n}\A_{-n}\A_{-n}\A_{-n}\;\;,\qquad  &\A_{-(n+1)}\A_{-n}\A_{-n}\A_{-n}\;\;, \cr
&\A_{-(n+1)}\A_{-(n+1)}\A_{-n}\A_{-n}\;\;,\qquad &\A_{-(n+1)}\A_{-(n+1)}\A_{-(n+1)}\A_{-n}\cr}
\eqlabel\exquad$$
The generalization (\parexcl) is then immediate.

\subsec{A quasi-particle reinterpretation of the effective  parafermionic exclusion principle}

Forbidding the states (\parexcl) is  equivalent to implementing an {\it effective 
$\ZZ_k$ parafermionic exclusion principle}. This interpretation  is made more precise in the present
subsection.

Let us come back to the Pauli exclusion principle:  two fermions cannot be put in the same state and this
translates into the statement $b_{-\nu} b_{-\nu} =0$.  The natural generalization is obvious: we cannot put
$k$ $\ZZ_k$-parafermions in the same state. This however lacks precision: what is the right meaning of `the
same state'? The state characterization should refer to the type of excitation. But this can be labeled
by the complete mode index of the creation operator (i.e., including its fractional part), -- which is minus
the conformal dimension,  referred to here as the energy --, or by its integral part.   But  `the same
state' characterization cannot pertain to the energy since the energy label always increases along a string
(e.g., the fractional part increases steadily by steps of $2/k$ from right to left). Hence, it has to refer to
the integral part of the mode. But the exclusion rule itself must also be related to  the energy of the
states.  In that regard, the energy of a state of $n$ identical excitations will refer to the {\it average
energy}.

We thus formulate the following `principle': in the $\ZZ_k$ model, we cannot add $p$ $\A_{-n}$ excitations to a
state that already contains
$(k-p)$ quasi-particles of type $\A_{-m}$,  with $n\geq m$, unless the average energy of the $n$ modes is
strictly greater than that of the
$m$ modes.  We now show
 that this  forbids of the $k$-strings listed in (\parexcl).
Consider the string
$$(\A_{-{n}})^p(\A_{-{m}})^{k-p}= A_{-{\nu_1}}\cdots A_{-{\nu_p}}
A_{-{\nu_{p+1}}}\cdots A_{-{\nu_k}}\qquad (\nu_i=n_i-f_i)\eqlabel\dede$$
acting on a state of ahcrge $q$. The average energy of the $\A_{-{n}}$ and the $\A_{-{n}}$ modes is respectively
$$
{\cal E}_n= n-2+{p-q\over k}\;,\qquad\quad {\cal E}_m= m-1+{p-q\over k}\eq$$so that ${\cal E}_n>{\cal E}_m$ implies $n> m+1$,  and this is equivalent to the exclusion of the complete set of
$k$-strings listed in (\parexcl).

This of course is not a proof of the selection rule (\parexcl) but  simply a  reformulation in
a quasi-particle language.

\subsec{On the `effective' nature of the exclusion principle and its $\ZZ_k$ origin}

%## formulation modifiee..et j,ai verifie
%## $\A_{-{3}}\A_{-{2}}\A_{{-2}}|0\R\not=0$

For $k>2$, the $\ZZ_k$ `effective exclusion principle' (\parexcl) cannot be a genuine
exclusion principle as in the
$k=2$ case. For this, it would be necessary that at least one  $k$-string (\parexcl)  be
exactly zero in analogy with the
$k=2$ case, where one of the two sequences (\exdeu) vanishes in view of the fermionic nature
of the $\A$ modes. It is simple to check, through examples, that the action of these $k$-strings is generically
not equal to zero.  For instance, for $k=3$,  we can check that
$\A_{-{2}}\A_{-{2}}\A_{-{1}}|0\R\not=0$ and $\A_{-{3}}\A_{-{2}}\A_{{-2}}|0\R\not=0$. The  meaning
of the restriction rule is confined to a {\it restriction on the  counting} of states: all states
containing
$k$-strings of the type (\parexcl) can be reexpressed in terms of other states which are free of them.   

Now, what is the origin of this restriction? It is bound to be rooted in the mere $\ZZ_k$ invariance,
namely the relation $(\psi_1)^k\sim \II$. Up to this point this condition has not been taken into account
properly, having been invoked only to argue that the charge of the descendent states is defined modulo $2k$.
But this condition means that at every level, {\it one needs to take out one state in each $k$-string}. This
is precisely what the restriction rule (\parexcl) does for us. That these conditions imposed on an ordered sequences indeed provide a basis, is demonstrated in
appendix C.

\newsec{Illustrative examples}

In this section, we display the states at the first few levels in some simple vacuum modules for $k=2$ and $3$. But before, we clarify the relation between the conformal dimension of a string of $\A$
operators and the sum of its integer modes.  

\subsec{The fractional dimension of a $p$-string}

Consider the module of $|\vph_\ell\R$ of relative charge $2r$. The various states all contain $m=kj+r$
$\A$ operators acting on $|\vph_\ell\R$. The level $s$ appropriate to a given sequence is obtained
by summing the mode indices (with reversed sign) and subtracting the contribution of the sum of the omitted
fractional parts. Explicitly, the $A$-version of the string reads
$$A_{-n_1+(1+2(m-1)+\ell)/k}\cdots A_{-n_m+(1+\ell)/k} | \vph_\ell\R\eq$$
The fractional parts $f_i$ add up to $$F=\sum_{i=1}^{kj+r} f_i= kj^2+2jr+j\ell+{r^2\over k}+{\ell r\over
k}\eq$$
The dimension of the string is then $\sum n_i-F$.  In the $r=0$ charge sector, this dimension is integer.
The character is constructed by summing the number of states at each level $s$ and by multiplying the result
by $q^{h_\ell-c/24}$, with $c=2(k-1)/(k+2)$. In a charged sector, the dimension of the string is not integral.
But if we redefine
$ s$ as
$$s'=s+{r^2\over k}+{\ell r\over k} -r\eqlabel\defsp$$
$s'$ is integer and the shift amounts to modify the prefactor from $q^{h_\ell-c/24}$ to $q^{h_\ell^{(r)}-c/24}$
where $h_\ell^{(r)}$ stands for the dimension of the basic field $\varphi_\ell^{(r)}$, associated to the
highest-weight state in the charge $2r$ sector:
$$h_\ell^{(r)}= h_{\ell}+{r(k-\ell-r)\over k}\eqlabel\stri$$

Summarizing, the states at level $s'$ in the module of $|\vph_\ell\R$ of relative charge $2r$, are generated
by the decompositions of the integer $n$ related to $s'$ by
$$n=\sum_{i=1}^{jk+r}n_i= s'+kj^2+j\ell+(2j+1)r\eqlabel\conte$$ into $kj+r$ parts subject to the
selection rules (\ferbass) and (\parexcl). In the following, the states will be represented by the
corresponding reversed  partitions, that is,
$$\A_{-n_1}\A_{-n_2}\cdots \A_{-n_m}|\varphi_\ell\R\qquad \rw \qquad 
(n_m,\cdots n_1)\eqlabel\ferbaspe$$ 

\subsec{The $k=2$ case}

Consider first the uncharged\foot{The use of the word `charge' in the context  of the Ising model is of
course formal since there is no physical notion of charge when $k=2$.} vacuum ($\ell=r=0$) module of the Ising
model.  All the states are then constructed by the application of an even number of ordered $\A$ modes,
forbidding pairs whose mode index do not differ by at most 2.  In other words, we consider partitions into an
even number of parts that differ by at least 2. The vacuum itself is represented by the void  partition 
and there is no state at level 1. The list of states at the next first few levels is 
$$\eqalignD{
&s=2: \; &(13)\cr
&s=3: \; &(14) \cr
&s=4: \; &(15)\;\;(24)\cr 
&s=5: \; &(16)\;\;(25)\cr
&s=6: \; &(17)\;\;(26)\;\;(35)\cr
&s=7: \; &(18)\;\;(27)\;\;(36)\cr
&s=8: \;&(19)\;\;(28)\;\;(37)\;\;(46)\;\; (1357)\cr
&s=9: \; &(1,10)\;\;(29)\;\;(38)\;\;(47)\;\; (1358)\cr
&s=10: \; &(1,11)\;\;(2,10)\;\;(39)\;\;(48)\;\;(57)\;\; (1359)\;\;(1368)\cr
}\eq$$
It is clear that the minimum level at which a string of length $2j$ occurs is
$2j^2$. 

The charged sector of the vacuum module ($\ell=0,\, r=1$) describes the Virasoro free-fermion module. It is
obtained from the vacuum by the application of an odd number of $\A$ operators. The first state is
$\A_{-1}|0\R$, of dimension 1/2. If we redefine the dimension of the other states with respect to this
highest state in the charged sector, we have then
$$\eqalignQ{
&s'=0: \; &(1)\quad\quad\quad &s'=4: \; &(5)\;\;(135)\cr
&s'=1: \; &(2) \quad &s'=5: \; &(6)\;\;(136)\cr
&s'=2:\; &(3)\quad &s'=6: \; &(7)\;\;(137)\;\;(246)\cr
&s'=3: \; &(4)\quad &s'=7: \; &(8)\;\;(138)\;\;(147)\;\;(146)\cr
}\eq$$

The remaining character is that of the spin field, which corresponds here to the case $\ell=1$. Its
charged and uncharged modules are equal, the $r=1$ case describing the disorder field. The partitions
associated to the first few states with $r=0$ are:
$$\eqalignQ{
&s=1: \; &(13)\quad\quad\quad &s=4: \; &(16)\;\;(25)\cr
&s=2: \; &(14) \quad\quad\quad &s=5: \; &(17)\;\;(26)\;\;(35)\cr
&s=3: \; &(15)\;\;(24)\quad\quad\quad &s=6: \; &(18)\;\;(27)\;\;(36)\;\; (1357)\cr
}\eq$$
Of course the partitions appearing here are identical to those already seen in the vacuum module. The only
difference is that the threshold level for the appearance of partitions of lengths $2j$ is shifted by a value
that depends upon $j$.

\subsec{The $k=3$ case}

For the uncharged vacuum module of the three-states Potts model, the list of the first few states obtained by
the application of $3j$-strings excluding the three 3-strings (\extri), is
$$\eqalignD{
&s=2: \; &(113)\cr
&s=3: \; &(114)\;\;(123) \cr
&s=4: \; &(115)\;\;(124)\;\;(133)\cr 
&s=5: \; &(116)\;\;(125)\;\;(134)\;\;(224)\cr
&s=6: \; &(117)\;\;(126)\;\;(135)\;\;(144)\;\;(225)\;\;(234)\;\; (113355)\cr
&s=7: \; &(118)\;\;(127)\;\;(136)\;\;(145)\;\;(226)\;\;(236)\;\;(244)\;\; (113356)\cr
&s=8: \; &(119)\;\;(128)\;\;(137)\;\;(146)\;\;(155)\;\;(227)\;\;(236)\;\;(245)\;\;(335)\cr
&~& (113357)\;\; (113366)\;\;
(113456)\cr}\eq$$
Up to level 14, the number of states in the 3-, 6-
and 9-  particle states are given by
$$\matrix{
s&0&1&2&3&4&5&6&7&8&9&10&11&12&13&14\cr
p{[3]}&~&~&1&2&3&4&6&7&9&11&13&15&18&20&23\cr
p{[6]}&~&~&~&~&~&~&1&1&3&5&9&13&21&28&41\cr
p{[9]}&~&~&~&~&~&~&~&~&~&~&~&~&~&~&1\cr}\eqlabel\decopa$$
where $p{[3j]}$ stands for the number of partitions of the integer $s+3j^2$ into $3j$ parts that forbid 
the subpartitions corresponding the 3-strings (\extri).
% \equiv p_{\Delta_3}^{[3j]}(s+3j^2)$).
The 9-particle sector starts then at level $s=14$, corresponding to the partition $(113355779)$. 
This counting of state can be compared to the sum of Virasoro characters
$\chi_0^{(3)}=q^{1/30}(\chi_{11}+\chi_{41})$ of the $(p',p)=(5,6)$ minimal model.

The two charged sectors ($r=1,2$) yield the modules of the two parafermionic fields (which are equivalent),  and
the counting of states reproduces that of the Virasoro field $\phi_{13}$. Similarly, with $\ell=1,2$, the
spin-field character $\chi_{23}$ is recovered and the charged sector yields the  energy-field module
$\chi_{21}+\chi_{31}$.

\newsec{Fermionic-type parafermionic character formulae}

\subsec{Character formulae in terms of restricted partitions}

The various states in the highest-weight module $|\vph_\ell\R$ of relative charge $2r$
are those obtained by the application of $r+jk$ $\A$ operators on the
highest-weight state, forbidding (\parexcl) and keeping track of the singular vector $|\chi_\ell\R$ in (\sig).
The counting of the resulting states level by level for a fixed value of the relative charge yields the
parafermionic highest-weight characters.

 Denote by $p_{\Dkl}^{[m]} (n)$ the number of partitions of $n= n_1+\cdots
+n_m$ into
$m=jk+r $ parts
 subject to the constraints:
$$\Dk n_i \equiv n_{i}-n_{i+k-1}\geq 2 \;,  \quad\quad n_i\geq n_{i+1}\geq 1\; ,\quad \quad
n_{m-k+\ell}\geq 2\eqlabel\dkde$$ Such partitions will be called $\Dkl$-{\it partitions}.

The $\vph_\ell$-character in the  relative charge sector $2r$ is thus (cf. section 4.1)
$$\chi_{\ell,r}= q^{h_\ell^{(r)}-c/24}\sum_{s'=0}^\y\sum_{j=0}^\y q^{s'} p_{\Dkl}^{[jk+r]}
(s'+j^2k+j\ell+(2j+1)r)\eqlabel\newca$$
where $h_\ell^{(r)}$ is given by (\stri). 
In particular, for the vacuum in the zero charge sector ($\ell=r=0$), it takes the simpler form
$$\chi_{0,0}= q^{-c/24}\sum_{s=0}^\y\sum_{j=0}^\y q^s p_{\Dkl}^{[jk]}
(s+j^2k)\eqlabel\newcava$$ 

Equation (\newca) is a truly remarkable result in that all the $\ZZ_k$ parafermionic characters, in all
possible charge sectors,  are written in terms of a single uniform expression.  In a fixed charged sector, all
the characters are determined by exactly the same restricted partitions, their level contribution being simply
shifted differently for different values of $\ell$.

 The expression (\newca) reflects the quasi-particle nature of the underlying basis in that it involves
no subtraction. However it is not yet written in a fermionic-sum representation. This last step is done in the
next subsection.

\subsec{Character formulae in fermionic-sum representation}

In order to obtain the fermionic form of the character (\newcava), we need to construct the generating
function for the $\Dkl$-partitions. The counting of such partitions at fixed
length can be reformulated into the counting of the number of solutions of simple Diophantine inequalities. 
The construction of the generating functions for such problems is well-known (see for instance [\ref{P.
MacMahon, {\it Combinatory analysis}, 2 vols (1917,1918), 3rd ed. reprinted by Chelsea,
1984.}\refname\Mac]) and it is used in appendix D to build up the
$k=2$ product form of the character  (\newcava) and that of the 3-particle sector of the
Potts vacuum character.  However, this method gets quickly cumbersome. 

Fortunately, the solution of this precise general problem has already been obtained by
Andrews [\And] (see also [\Andr]) and in this
section we use directly that result. The number of
$\Dkl$-partitions of $n$ into $m$ parts that contain at most $k-\ell$ parts equal to 1 is given by (cf.
Theorem 1 of [\ref{G.E. Andrews, Houston J.
Math. {\bf 7} (1981) 11.}\refname\Andrr] with $d=0$ and  $i=k-\ell+1$ -- or eq. (7.38) of [\Andr])
$$\sum_{n,m=0}^\y p_{\Dkl}^{[m]}(n) \, q^nz^m= \sum_{m_1,\cdots,m_{k-1}=0}^\y { q^{N_1^2+\cdots+
N_{k-1}^2+L_{k-\ell+1}} z^{N} \over (q)_{n_1}\cdots (q)_{n_{k-1}} }\eqlabel\ande$$
where $$\eqalignD { & N= N_1+\cdots+N_{k-1}\, , \qquad &N_i= m_i+\cdots +m_{k-1}\cr & L_j=N_j+\cdots N_{k-1}\,,
& L_k=L_{k+1}=0 \cr}\eq$$
%(q)_n= (1-q)\cdots (1-q^n) ($L_k=0$).

Let us apply this to the general case of a charged character of $\vph_\ell$. Recall that the partitions we are
interested in are related to the level
$s'$ (itself related to $s$ by (\defsp)) via  
$n=s'+j^2k+(2j+1)r$. In order to count 
$jk+r$-string partitions, we set $z^m=0$ if $m$ is not equal to $r$ mod $k$.  
For $m=jk+r$, we will fix $z^m$ in order to adjust the power of $q$ in our sum over $n$ to that of the level
$s$, namely, to transform
$$\sum_{n} p_{\Dkl}^{[jk+r]}(s'+j^2k+j\ell+(2j+1)r) \, q^nz^m \;\rw \;\sum_{s} p_{\Dkl}^{[jk+r]}(s'+j^2k+j\ell+(2j+1)r) \,
q^{s'}\eq$$
For this we need to set
$$ z^m=\left\{ \matrix{ q^{-(m-r)(m+r+\ell)/k -r}\qquad& {\rm
when}\quad m=jk +r \cr 0 &{\rm otherwise}\cr} \right.\eq$$ 
The identity (\ande) becomes then
$$\eqalign{ &\sum_{s',j=0}^\y p_{\Dkl}^{[jk+r]}(s'+j^2k+j\ell+(2j+1)r) \, q^{s'} \cr
&\qquad \quad = 
\sum_{m_1,\cdots,m_{k-1}=0 \atop 
N=r\;{\rm mod}\;k}^\y { q^{N_1^2+\cdots+ N_{k-1}^2- (N-r)(N+r+\ell)/k-r+L_{k-\ell+1}}  \over
(q)_{m_1}\cdots (q)_{m_{k-1}} } \cr}\eqlabel\lepoo$$
Note that on the rhs, the contribution to the $jk$-particle sector is selected by the
condition   $N = \sum im_i=jk$.
Multiplied by $q^{h_\ell^{(r)}-c/24}$, this yields the general form of the $\ZZ_k$ charged characters in
a fermionic-sum representation. With this factor, this is precisely the
Lepowsky-Primc formula (theorem 9.4  of [\LP]).

\newsec{Conclusion}

We have presented a new  basis for the space of states in $\ZZ_k$ parafermionic conformal field theories. It
yields a `faithful' description of the states in irreducible modules, faithful in the sense that it
does not require the explicit  subtraction of an infinite number of singular-vectors.\foot{The existence of
such a basis was not  expected in the initial stage of our study of parafermionic theories - cf. the
discussion following eq (1.1) of [\JM].} The basis is formulated in terms of a single type of modes, namely,
the modes $\A$ of the parafermionic field
$\psi_1$. States in a highest-weight module are thus of the form $\A_{-n_1}\cdots \A_{-n_m} |\varphi_\ell\R$.
The various constraints on the mode-label  $n_i$ 
implement in a very natural way:

\n (1) - an ordering: $n_i\geq n_{i+1}$;

\n (2) - the highest-weight nature of $|\varphi_\ell\R$: $n_i\geq 1$; 

\n (3) - the subtraction of one `primary' singular vector, the one that can be
formulated solely in terms of the
$\A$ modes -- cf. $|\chi_\ell\R$ in (\sig) -- and this is taken into account by the `boundary condition':
$n_{p-k+\ell}\geq 2$;

%## legere modif de (3)

\n (4) - the $\ZZ_k$ invariance, which implies that $(\psi_1)^k\sim \II$, which is taken care of by the
forbidding of the $k$-strings (\parexcl), or equivalently, the $\Dk$ condition: $n_i-n_{i+k-1}\geq 2$. 
As explained in section 3, this condition can be formulated in terms of an effective exclusion principle.

 The counting of partitions of $n= n_1+\cdots + n_m$ subject to these conditions (called $\Dkl$-partitions)
leads directly to the $\ZZ_k$ characters written in a manifestly positive series expansion in $q$ (cf. eq.
(\newca)). 
 Using Andrews' identity (\ande), the generating function of these restricted partitions can in turn be written
as sums of products.  The resulting character formula (\lepoo) is precisely the one obtained by Lepowsky-Primc
[\LP].

Since a character expression is the transcription of a basis, our new fermionic-type basis is bound to be
equivalent to that of [\LP]. Actually, our last step (namely, 
section 5.2)
is  identical to the final step in the derivation of the Lepowsky-Primc formula: we count the same type
of objects since at this point,  {\it our
basis and theirs turn out to be equivalent}. The precise equivalence is obtained when their
$z_\a(-n)$ operators are replaced by our $\A_{-n}$ ones (cf. their theorem 6.8 and the constraints
(6.13)).\foot{We stress in that regard that the transition from fractional to
integer modes for writing of strings of parafermionic modes clarified the way for the formulation of the
standard basis in [\JM] and also the fermionic-type basis presented here; and it is with this notation that the
connection with [\LP] becomes crystal clear.} However,  their approach, which is  Lie-algebraic, is 
completely different from ours. Their original aim was to obtain representations of standard modules in
$\suh(2)_k$ by constructing $z_\a$ operators in the so-called homogeneous picture, meaning that they
are forced to commute with those of the homogeneous Heisenberg algebra, itself simply a 
$\uh(1)$ algebra. Given that the $\ZZ_k$ parafermionic theory has a natural description in terms of the coset
$\suh(2)_k/\uh(1)$, it is clear, {\it a posteriori}, that the $z_\a$ operators have to be directly related to
the parafermionic fields. The present results give the precise correspondence mentioned above. 

%** ajout
Our basis is also equivalent to the $\ZZ_k$ quasi-particle basis displayed in [\bouschou], 
but presented there in
a more complicated-looking form (cf. their eqs (4.3)-(4.5)), which is further valid only for the vacuum,  and
given without proof -- apart from the plain fact that it yields the right character.\foot{Their second vacuum
basis (their eq. (4.10)) appears to be a mere rearrangement of the first one and it is not proved either.}

Although at the end, we reproduce the Lepowsky-Primc basis in a disguised form, we stress that
there is a great interest for a purely parafermionic derivation of this fermionic-type basis. 
Firstly, the parafermionic point of view is clearly more physical, having a natural
quasi-particle interpretation. In  this spirit, we have seen that the $\Dk$ condition comes from our
effective $\ZZ_k$ exclusion principle, which in turn reflects the built-in
$\ZZ_k$ invariance, through $(\psi_1)^k\sim \II$.  Secondly, there exist generalized
parafermionic theories that do not have a simple coset description of the type $\gh/\uh(1)^r$ -- to which the
Lepowsky-Primc construction should be limited. This is certainly the case for the $\ZZ_k^{(\beta)}$  models in
which  the dimension of the parafermionic field $\psi_n$ reads
$\beta n(k-n)/k$, with $\beta$ integer - cf. appendix A of [\ZFa]). The search for
a quasi-particle basis in those cases requires purely parafermionic techniques such as the ones considered
here. \foot{In that vein, it should be noticed that in a parafermionic approach, it is somewhat natural to
first unravel the bosonic form of the character given that it contains some information (namely one
`primary' singular vector) needed in the formulation of the quasi-particle basis.}

%##: ajout de foot ci-haut

On the other hand, we stress that the quasi-particle basis displayed here is not the unique quasi-particle
basis underlying $\ZZ_k$ parafermionic conformal theories. Indeed, as stressed in [\KKMM], there are two
different integrable models whose scaling limit is the $\ZZ_k$ parafermionic theories: the  $su(k+2)$
RSOS model at the boundary of regimes I/II and the $\ZZ_k$ spin chains [\ref{V.A. Fateev and
A.B.Zamolodchikov, Phys. Lett. {\bf 92A} (1982) 37.}] and these have different quasi-particles description. 
Actually, a second fermionic sum representation has been found in [\DKKMM, \KMM]. It would be interesting to
unravel the structure of the underlying basis. In that vein, it would be of interest to connect each fermionic
sum to a particular integrable perturbation by an energy operator [\ref{V.A. Fateev, Int J. Mod Phys. {\bf
A6} (1991) 2109.}].

We
conclude with some general comments.  The first one pertains to the Andrews-Gordon-Rogers-Ramanujan-Schur
identities themselves.  The classical  Rogers-Ramanujan-Schur relations (\yang) -- which correspond to
(\gordon) for
$k=2$ -- have the following combinatorial interpretation (see e.g., [\Andr] corollaries 7.6 and 7.7): the
number of partitions of
$n$ into
$m$ numbers such that $ n_i-n_{i+1}\geq 2$, with the `boundary condition' $n_m\geq 1+a$, is
equal  to the number of partitions of $n$ into parts not congruent to $0,\pm (2-a)$ mod 5. The most immediate
generalization that one could think of is to replace the 2 in the difference condition by a larger integer.
Surprisingly, that faces various no-go theorems (cf. the introduction of [\Gor]). The breakthrough for
generalizations came when Gordon considered the Glaisher's $k$-periodicity condition, i.e., a condition on the
difference $n_i-n_{i+k-1}$, together with the lower bound 2. The analysis presented here allows us to state
that one side of the combinatorics underlying these classical  Rogers-Ramanujan-Schur  identities has a very
natural fermionic field-theoretical interpretation. In that vein, the corresponding side of the Andrews-Gordon
generalization similarly has a quite natural parafermionic 
interpretation\foot{The `boundary' part of this statement should be slightly
qualified, however. The Rogers-Ramanujan-Schur `boundary condition' is not exactly the one which is appropriate
to the  Ising model and this generalizes the higher $k$ cases. Both (combinatorial and parafermionic)
conditions would be identical if the singular vector
$|\chi_\ell\R$ was given by $\A_{-1}^{k-\ell} |\vph_\ell\R$ instead of  $\A_{-1}^{k-\ell+1} |\vph_\ell\R$.} in that it constrains $k$ adjacent
parts along a sort of $\ZZ_k$ exclusion principle.\foot{We found afterwards that similar comments are presented in [\ref{J. Lepowsky and R.L. Wilson, Proc. Nat. Acad. Sci. USA {\bf 78} (1981) 7254.}\refname\LW], where this observation appears as a motivation for studying $Z$-algebras. See also the introduction of [\ref{C. Dong and J. Lepowsky {\it Generalized vertex algebras and relative vertex operators}, Birkha\"user, 1993.}\refname\DL]. In the last chapter of this book, the authors also give the correspondence between the $Z$-algebras and the $\ZZ_k$ parafermionic theory.}  

%## foot ci-haut

The pervasive nature of the $\Dk$ condition is quite impressive. In
conformal field theory, this ubiquity is itself signaled in the  many occurrences of fermionic-type character
formulae that have appeared recently and whose core is precisely this $\Dk$ condition. And this condition
reflects a $\ZZ_k$ invariance. It might be a hint that parafermions could provide the core description of {\it
a priori}  quite unrelated models. In particular, it is
rather remarkable to see that the characters of the
$(2,2k+1)$ minimal models are also of the Lepowsky-Primc form [\FNO], which strongly suggests an
underlying parafermionic construction.

\appendix{A}{The bosonic-type $\ZZ_k$ character formula}

The character of the highest-weight module $\vph_\ell$ with relative charge $2t\geq 0$\foot{Recall that in view
of the field identifications, it suffices to consider the non-negative charge sector only.} is  given by [\JM]
$$\chi_{\ell,t}= q^{h_\ell^{(t)} -c/24-t }\sum_{s=0}q^s
g_{\ell,t}(s)\eqlabel\genecara$$ where 
$
h_\ell^{(t)}$ is given by (\stri)
and $g_{\ell,t}(s)$ corresponds
to  the number of states at level $s$ having charge $2t$ and it is given by $$g_{\ell,t}(s)
= \sum_{p=0}^\y g_{\ell,t,p}(s)\eqlabel\nbi$$ with
$$g_{\ell,t,p}(s) = G_{0,t}\delta_{p,0} -G_{r_{p-1},s_p+t}
- G_{r'_{p},s'_{p-1}+t}+G_{r_{p},s_p+t} +G_{r'_{p},s'_p+t}
\eqlabel\nbu$$  
where $$
G_{a,b}(s)\equiv\sum_{j=0}^s\sum_{s_1=j}^s 
p^{(j-a)}(s_1-j)\,p^{(j-b)}(s-s_1)\eqlabel\nbb$$
with $a,b$ two non-negative integers. 
The indices of $G$ in (\nbu) are defined as follows:
$$\eqalignD{ 
r_{p}&= (p+1)[(p+1)(k+2)+\ell+1]\;,\qquad &r'_{p} = (p+1))[(p+1)(k+2)-\ell-1]\cr 
s_p &= (p+1)[(p(k+2)+\ell+1]\;,\qquad 
&s'_p =(p+1)[(p+2)(k+2)-\ell-1]\cr }\eqlabel\sytb$$
In (\nbb), $p^{(j-a)}(n)$ stands for the partition of $n$ into {\it at most} $j$ parts\foot{For instance, $
p^{(3)}(6)= 7$ since 6 can be decomposed in 7 different ways in sums of at most three integers:
$6=5+1=4+2=3+3=4+1+1=3+2+1=3+1+1=2+2+2$.} and
$$ p^{(0)}(n)= \delta_{n,0} \;\; , \qquad \quad
 p^{(j\geq 0)}(0)=1\;\; , \qquad \quad  p^{(j<0)}(n)=0 \eqlabel\conve$$
which implies in particular that
$G_{a,b}(s)= 0$ if $s<\,{\rm max}\,(a,b)$.  Simple illustrations of this formula can be found in section 6 of
[\JM].

These character formulae  can be written somewhat more compactly in terms of the generating function for the
factors $\sum q^s G_{a,b}(s)$. Notice that
$$\sum_{s=0}^\y q^s G_{a,b}(s) = \sum_{s=0}^\y\sum_{j=0}^s\sum_{s_1=j}^s 
 q^sp^{(j-a)}(s_1-j)\,p^{(j-b)}(s-s_1)\eq$$
With the conditions (\conve) and $p^{(j)}(n<0)=0$, we can start the sum over $s_1$ from 0, extend that of $j$
to $\y$ and reexpress that over $s=s_1+s_2$ as one over $s_2$, with the result:
$$
\sum_{s=0}^\y q^s G_{a,b}(s)  = \sum_{j=0}^\y \left(\sum_{s_1=0}^\y 
q^{s_1} p^{(j-a)}(s_1-j)\right)\left(\sum_{s_2=0}^\y q^{s_2}p^{(j-b)}(s_2)\right)
 = \sum_{j=0}^\y {q^j\over (q)_{j-a}(q)_{j-b}} \eqlabel\parfait$$
The occurrence of the $(q)_n$ factors  makes  these characters sorts of hybrid between bosonic and fermionic
forms.

\appendix{B}{Bosonic vs fermionic characters  of the $\ZZ_\y$
model}

In the $k\rw\y$ limit, the parafermionic theory reduces to a theory of two independent complex bosons [\ref{P.
Jacob and P. Mathieu, {\it Parafermionic Jacobi identities,  Sugawara construction and generalized commutation
relations}, in preparation.}\refname\JMc]. In that limit, the bosonic form of the parafermionic character
becomes rather simple in that there is a single contributing singular vector, namely $(\A_0^\da)^{\ell+1}\,
|\vph_\ell\R$. 

Let us consider for simplicity the
vacuum module of zero relative charge, in which case the single contributing singular vector arises at level
1. The $r=0$ vacuum character reduces then to (with $c=2$):
$$\chi_{0,0}= q^{-1/12}\sum_{s=0}^\y q^s \,[G_{0,0}(s)-G_{0,1}(s)]\eqlabel\ddr$$

This will be shown to be equivalent to the character of the vacuum module in a theory of two
independent complex bosons, with modes $a_n$ and $\ab_n$, when evaluated in the zero-charge sector. The 
highest-weight conditions being
$a_{n}|0\R=
\ab_n|0\R=0$ for $n\geq 0$, the different states at level $s$ in the uncharged vacuum module are of the form
$$a_{-n_1}\cdots a_{-n_j}\ab_{-m_1}\cdots \ab_{-m_j}|0\R \qquad{\rm with}\quad  n_i\geq
n_{i+1}\, ,\qquad m_i\geq m_{i+1}\eq$$   and
$$\sum_{i=1}^jn_i=s_1\qquad \sum_{i=1}^jm_i=s-s_1\eq$$
The corresponding character is simply
$$\chi_0^{{\rm cb}}=  q^{-1/12}\sum_{s=0}^\y q^s \sum_{j=0}^\y \sum_{s_1=j}^\y
p^{[j]}(s_1)p^{[j]}(s-s_1)\eqlabel\cplx$$
where, as usual, $p^{[j]}(n)$ stands  for the number of partitions of $n$ into {\it  exactly} $j$ parts.

Let us now demonstrate that (\ddr) reduces to (\cplx).  For this we simply note that
$$\eqalign{
G_{0,0}(s)-G_{0,1}(s) &= \sum_{j=0}^\y \sum_{s_1=j}^\y
p^{(j)}(s_1-j)[ p^{(j)}(s-s_1)- p^{(j-1)}(s-s_1) ]\cr
&= \sum_{j=0}^\y \sum_{s_1=j}^\y
p^{(j)}(s_1-j) p^{[j]}(s-s_1) \cr
&= \sum_{j=0}^\y \sum_{s_1=j}^\y
p^{[j]}(s_1)p^{[j]}(s-s_1) \cr}\eq$$
Since $p^{(j)}(n)$ is the number of partitions of $n$ into {\it  at
most} $j$ parts, it is  clear that $p^{(j)}(n)-p^{(j-1)}(n)=p^{[j]}(n) $. In the last
line, we used the relation
$p^{(j)}(n-j)=p^{[j]}(n)$.\foot{The respective generating functions of these partitions are
$$\sum_{j=0}^\y p^{(j)}(n)q^n = {1\over (q)_j}\qquad\qquad  \sum_{j=0}^\y p^{[j]}(n)q^n = {q^j\over (q)_j}$$
from which the above identity follows directly.}  The 
number of states at level
$s$ is thus the same as in the vacuum module of the two-complex-boson theory.

The character (\cplx) can easily be put in a fermionic-type form.  The generating function of a
free-boson character is simply $q^{-1/24}(q)_\y^{-1}$. To take the charge into account, we introduce a
parameter $z$ into the Euler function:
$$(z;q)_\y\equiv \prod_{n=1}^\y (1-zq^n)\eq$$ States in the product Fock space of two complex bosons are thus
generated by the product $q^{-1/12} [ (z;q)_\y (z^{-1};q)_\y ]^{-1}$. To get the complete set of 
states of zero charge, we project this expression onto the $z^0$ sector. Using the Euler relation 
(cf. theorem 2.1 of [\Andr] with $a=0$ and $t=zq$):
$${1\over (z;q)_\y}= \sum_{j=0}^\y {(zq)^j\over (q)_j}\eq$$
it ready follows that 
$$\left.{1\over (z;q)_\y \, (z^{-1};q)_\y}\right|_{z^0}= \sum_{j=0}^\y {(q)^{2j}\over (q)_j^2}\eq$$
so that 
$$\chi_0^{{\rm cb}}=  q^{-1/12}\sum_{j=0}^\y {(q)^{2j}\over (q)_j^2}\eqlabel\clxb$$
Of course, the above sum is nothing but (cf. (\parfait)):
$$ \sum_{s=0}^\y q^s [G_{0,0}(s)-G_{0,1}(s)]=\left({q^j\over (q)_{j}(q)_{j} }-{q^j\over
(q)_{j}(q)_{j-1} }\right)\eq$$

The  expression (\clxb) is rather different from  (\lepoo) evaluated in the $k\rw \y$ limit and we have not made
the direct correspondence. Note however that there are many ways that even the Euler function $(q)_\y^{-1}$ for
instance  can be transformed into a multiple $q$-series (cf. theorem 3 of 
[\Andrr]):
$${1\over (q)_\y} = \sum_{m_1,\cdots , m_{k-1}}{\ q^{N_1^2+\cdots
 N_{k-1}^2}\over (q)_{m_1}\cdots (q)_{m_{k-2}} (q)^2_{m_{k-1}} }\eqlabel\andrews$$
In this identity, the number $k$ is totally arbitrary (and with $k=2$, this reduces to (\bosca)).  

It is also
interesting to recall the form of the vacuum string function in that limit [\KP,
\DQ,
\Nem]: 
$$b_0^{(\y)} =  q^{-1/12}{1\over (q)_\y^2}\left [ 1+ 2\sum_{p=1}^\y (-1)^p q^{p(p+1)/2}\right]\eqlabel\sift$$
The  equivalence of this expression with  (\cplx) provides a nice identity.

\appendix{C}{From the $\ZZ_k$ invariance to the restriction rules}

In this appendix, we prove that, as a result of the model's $\ZZ_k$ invariance, the $k$ distinct $k$-strings
(\parexcl) can be forbidden.  Before plunging into the details, we outline  the argument.

At first, we  rephrase the condition 
 $(\psi_1)^k\sim \II$ in terms of modes of the $k$-th composite parafermionic field $\psi_k$ and argue that
this implies one linear relation at each level. This condition is then reexpressed in an
ordered basis formulated solely in terms of the $\A$ modes.  We show that this naturally amounts to forbid the
product of
$k$ strings $\A_{-n_1}\cdots \A_{-n_k}$ that satisfy $n_k+1\geq n_1$. This inequality selects precisely the
$k$-strings (\parexcl) and this set indeed generates one forbidden state at each level in product of $k$ $\A$
descendant operators. 

In a second step, the compatibility requirements of these selection rules imposed
within strings of length $m>k$ are studied.  At this level, we must ensure that there are no induced
contradictions and that exactly the right number of constraints is still generated.  Arguing from that
perspective shows that the selection rules (\parexcl) are singled out in that they forbid sequences of
states with indices `as equal as possible'.

% \subsec{Restrictions on $k$ strings}

We thus start by displaying the commutation relations between the modes of the parafermions $\psi_r$ and
$\psi_s$, whose OPE reads
$$\psi_r (z) \,\psi_{s} (w) \sim {c_{r,s}\over  (z-w)^{2rs/k}}\;
 \psi_{r+s} (w)  \qquad (r+s\leq k)\eq$$
with 
$$c^2_{r,s}={ (r+s)!  (k-r)!  (k-s)! \over 
 r! s!  (k-r-s)! k!}\eq$$
With $\psi_r$ expanded in modes as
$$ \psi_r(z)\phi_{q}(0)  = \sum_{m=-\y}^\y
z^{-rq/k-m-r}A^{(r)}_{r(r+q)/k+m}\, \phi_{q}(0)\eqlabel\modep$$
where $\phi_q$ is an arbitrary field of charge $q$,  
the mode commutation relation between
$A^{(r)}_u$ and $A^{(s)}_v$ reads
$$ \sum_{l=0}^{\infty} C^{(l)}_{2rs/k-1} \left[ \A^{(r)}_{n-l-r}
\A_{m+l-s+1}^{(s)} + \A_{m-l-s}^{(s)} \A^{(r)}_{n+l-r+1}
 \right]  =c_{r,s}\,\A^{(r+s)}_{n+m-r-s+1} 
\eqlabel\comma$$
In this last expression, the field $\phi_q$ and the fractional parts of the modes have been omitted; $C^{(l)}_t$ is defined
after (\combsi).

Since $(\psi_1)^k\sim \II$, the modes $\A^{(k)}_p$ are in fact simply  delta functions. Being interested in
 relations between states, we thus write $\A^{(k)}_p\sim 0$.
That yields one condition for every value of $p$.  By setting $r=k-1$, $s=1$, $m=-n_k$ and
$n=-n_1-\cdots -n_{k-1}+k-1$ in (\comma), we find
$$ \sum_{l=0}^{\infty} C^{(l)}_{1-2/k}  \A^{(k-1)}_{-n_1-\cdots -n_{k-1}-l}
\A_{-n_k+l} \sim - \sum_{l=0}^{\infty} C^{(l)}_{1-2/k} \A_{-n_k-
l-1} \A^{(k-1)}_{-n_1-\cdots -n_{k-1}+l+1}   
\eqlabel\commaa$$
We now use (\comma) recursively to eliminate the modes of the composite parafermions $\A^{(k-i)}$
starting from $i=1$ up to $i=k-2$.  
Note that the precise manner in
which
$\A^{(k)}_n$ is first decomposed into two parts
$\A^{(k_1)}_{n-m}\A^{(k_2)}_m$ with
$k_1+k_2=k$ (here $k_1,\,k_2$ being chosen to be 1 and $k-1$), and then in which the  various $\A^{(k-i)}_n$
are further split, does not influence the final result in view of  associativity. In other words, the
different ways of separating $\A^{(k)}_n$ do not provide different relations.

% In the first step, we use  (\comma) 
% to express the terms on the two sides 
% of (\commaa) as  sums over products 
% of $k$ $\A$ factors.

% We now treat the general case.   
The
splitting of the modes in the initial decomposition (\commaa)
 has been chosen in such a way that the lhs can
produce the term
$\A_{-n_1}\A_{-n_2}\cdots
\A_{-n_k}$ (using (\comma) recursively) for which we assume that the indices
$n_i$'s are ordered: $n_1\geq n_2\geq \cdots \geq n_k$.  Let us isolate this term on the lhs.  It
can thus be expressed as a linear combination of the form
$\sum c_{1\cdots k} \A_{-n'_1}\A_{-n'_2}\cdots
\A_{-n'_k}$, with $\sum n_i= \sum n'_i$. These terms are supposed to have been  reordered after
the decomposition. Now all these ordered terms are necessarily independent of $\A_{-n_1}\A_{-n_2}\cdots
\A_{-n_k}$ if it happens that the negative of their first mode index is strictly greater than that of
$\A_{-n_1}\A_{-n_2}\cdots \A_{-n_k}$, i.e., than $n'_1>n_1$. Now if $n_1'=n_1$, it is clear that 
$\A_{-n_1}\A_{-n'_2}\cdots
\A_{-n'_k}$ is not necessarily independent of our isolated term and for this, we obviously need to
compare the index with smallest $i$ such that $n_i'\not=n_i$.  

 Enforcing  that minus the first
index of every term on the rhs be greater or equal to
$n_1$, leads to a sequence of constraints which we now derive.
For this, we need to write down all possible terms which occur in the decomposition of the two sides of
(\commaa).  If the leftmost operator is already of the
$\A$ type, no further analysis is required. This is the case for the rhs of (\commaa). On the lhs, let us
redefine
$l\rw l_1$. Then, by using (\comma) once more, we get (do not caring about relative  coefficients)
$$\eqalign{
\A^{(k-1)}_{-n_1- \cdots - n_{k-1}-l_1}
 \sim ~&\A^{(k-2)}_{-n_1-\cdots - n_{k-2}-l_1-l_2} \A_{-n_{k-1}+l_{2}}\cr  
 +  &\A_{-n_{k-1}-l_{2}-1} \A^{(k-2)}_{-n_1-\cdots - n_{k-2}-l_1+l_2+1}\cr}\eq$$ 
The second term on the rhs has the desired form (an $\A$ factor at the left).  We then need to transform the
first one. It is clear that this first term, upon complete decomposition with similar index splitting, 
will generate the term
$\A_{-n_1}\A_{-n_2}\cdots
\A_{-n_k}$. Moreover, at each step of the process, there will be a term produced whose leftmost contribution
will be
$\A_{-n_{k-i}-l_{i+1}-1}$. Recall that it is the mode index of all these leftmost terms that need to be
compared with
$n_1$. This leads to the following list of constraints, in their most stringent version (i.e., with
$l_{i+1}=0$):
$$n_{i}+1 \geq n_1\eq$$
Since the $n_i$'s are supposed to be ordered, all these constraints are consequences of a single one, namely
$$n_{k}+1\geq n_1\eqlabel\conf$$

Quite remarkably, if all these constraints are
enforced and that
$n_1'=n_1$, then it follows that $n_i'=n_i$ for all other $i$.  $\A_{-n_1}\A_{-n_2}\cdots \A_{-n_k}$ itself is
thus recovered and it is transferred to the lhs.  This shows that when all the constraints are satisfied on the
first index of the prime terms, $\A_{-n_1}\A_{-n_2}\cdots \A_{-n_k}$ can be eliminated.

The condition (\conf) 
eliminates exactly one state at every level. Actually, it selects precisely the
$k$ terms listed in (\parexcl).

We now investigate the compatibility requirement associated to the elimination of the $k$-strings  (\parexcl)
when they are immersed within strings of length $m>k$.  Let us forget for a moment the special structure of
the $k$ strings that have been selected by the above analysis and focus on the plain fact that there is one
constraint at each level that results from the $\ZZ_k$ invariance.  Suppose that we select one such constraint
at each level  in some well-prescribed way. The point we want to stress here is that there is no guarantee that
these eliminated states, when considered within longer strings, will yield the correct number of constraints.

Take the $k=3$  uncharged vacuum module. At each level we have to eliminate one state. We could select the one with
largest value of $n_1$. At $s=3$, that would amount to eliminate $\A_{-4}\A_{-1}\A_{-1} $. Now consider the
$r=1$ vacuum module. At $s'=6$ we get the following states:
$$\eqalign{ 
s'=6: \quad &\A_{-5}\A_{-1}\A_{-1} \A_{-1} \quad \A_{-4}\A_{-2}\A_{-1}\A_{-1} \quad
\A_{-3}\A_{-3}\A_{-1}\A_{-1}\cr & \A_{-3}\A_{-2}\A_{-2}\A_{-1}\quad \A_{-2}\A_{-2}\A_{-2}\A_{-2}\cr}
\eqlabel\sixc$$ In this context, the elimination of the 3-string $\A_{-4}\A_{-1}\A_{-1}$ is problematic since we do not encounter the state $\A_{-2}\A_{-4}\A_{-1}\A_{-1}$. Hence,
not enough constraints are generated. 

The $k$ distinct $k$-strings that are eliminated in the
quasi-particle basis have the noteworthy structure of being precisely those products of
$k$ operators whose mode indices are `as equal as possible'. This has the important implication that within any
ordered sequence of operators, any one of the $k$ sequences (\parexcl) cannot appear separated, that is, no 
reordering is  required  for bringing together a sequence of a  $k$-string of the type (\parexcl).
In our example, it means that we eliminate $\A_{-2}\A_{-2}\A_{-2}$ at level 3 and this state is found either
in the form $(\A\A\A)\A$ or $\A(\A\A\A)$ in any ordered sequence of 4-strings.

Is it possible that too many constraints be generated?
Suppose that  we eliminate $\A_{-2}\A_{-2}\A_{-1}$ at $s=2$  and
$\A_{-3}\A_{-3}\A_{-1}$ at $s=4$.  In the $s'=6$ list (\sixc), we would then eliminate both
$\A_{-3}\A_{-3}\A_{-1}\A_{-1}$ and
$\A_{-3}\A_{-2}\A_{-2}\A_{-1}$. But a few lines computation shows that $\A_{-3}\A_{-2}\A_{-2}\A_{-1}\sim
\A_{-3}\A_{-3}\A_{-1}\A_{-1}$. That is, reshuffling the indices of an excluded state generates another
excluded state. Here it would thus suffice to eliminate only one of $\A_{-2}\A_{-2}\A_{-1}$ or
$\A_{-3}\A_{-3}\A_{-1}$ to eliminate the proper number of terms at level $s'=6$.This is a circumstance in which too many states are eliminated as a result of a
non-compatible choice of excluded 3-strings.
 
Now, that the states which are eliminated are those with indices `as equal as possible' guarantees the
compatibility of the elimination process itself. The transformation of such a state $\A_{-n_1}\A_{-n_2}\cdots
\A_{-n_m}$  into other ordered states implies a redistribution of the 
 integers
$(n_1, n_2,\cdots , n_m)$ into $(n'_1, n'_2,\cdots , n'_m)$. In this way, $n_m$ cannot be increased and the
new states are obtained by modifying the labels successively from right to left. Starting the process with a
state including a substring in the list (\parexcl), it is guaranteed that no state containing either this
substring or another substring in this list  can  be encountered later in the process. In other words, the
linear decomposition is triangular and thus manifestly coherent.

Let us finally comment on the linear dependence of the states (\ferbas), disregarding for the moment the restrictions (\restru). That the modes can be ordered in that way -- up to possible additional constrains -- is proved in [\JM]. Now, these ordered states are easily proved to be independent.  Consider first 2-strings and see if there exists some constants $a_{n_1,n_2}$ (not all zero) such that $\sum a_{n_1,n_2}\A_{-n_1}\A_{-n_2}=0$ for $n_1+n_1=N$ and $n_1\geq n_2\geq 1$. Using the commutation relations, the term with largest value of $n_1$ can be written as a linear combination of the other ordered states {\it plus} a number of non-ordered states (with computable non-zero finite coefficients). Hence, there are no $a_{n_1,n_2}$'s such that the above sum vanishes. For longer ordered strings, the argument is similar. Consider a subgroup of ordered $p$-strings at a given level and suppose that there is a unique term with largest value of $n_1$. Commuting the first two $\A$'s necessarily generates one non-ordered state whose second index is $-n_1$ (and the argument with 2-strings shows that it cannot vanish). Further commutations can only displace this $-n_1$ index along the string. Therefore there are terms that cannot be reordered and this implies the linear independence of such a set of ordered states. Now if in our subgroup of ordered strings, the state with largest value of $n_1$ is not unique, we repeat the argument with the index $n_j$ that orders the various strings with $n_i=n_i'\;, i=1,\cdots, j-1$ (e.g., $n_j>n_j'$). This completes the proof of the linear independence of the states (\ferbas).  Imposing the restriction (\restru) cannot spoil this linear independence: it only amounts to forbid one state per level.

\appendix{D}{Simple examples of generating functions for $\Dkl$-partitions from the
MacMahon method}

In this appendix, we construct the simplest generating functions 
for the $\Dkl$-partitions, mainly in 
order to get a more intuitive handle on this counting problem.    A
$\Dkl$-partition  of $n$ of length $p$ is characterized by the set of inequalities (\dkde).
Finding the generating function for all partitions $p_{\Dkl}^{[m]}(n)$ with $j$ and $k$ fixed is equivalent
to count the number of solutions to this system of inequalities.
This problem, in turn, can be tackled by 
the MacMahon algorithm for
constructing the generating function of the solutions of a system of linear Diophantine inequalities. 
For a detailed presentation,
see 
section VIII of vol. 2 in [\Mac].

To illustrate the method, let us first consider the simplest possible situation where $k=2$, $\ell=0$, $j=1$
and
$r=0$ (and write $\Delta_{k,0}=\Dk$).  These correspond to the partitions of $n=s+2$ in two numbers $n_1+n_2$
subject to the conditions
$$n_1\geq n_2+2, \qquad n_2\geq 1\eq$$ Consider then the function
$$G_1^{(2)}= {\la_1^{-2}\la_2^{-1}q^{-2}\over (1-\la_1 q)(1-\la_2 \la_1^{-1} q)}= \sum_{n_1,n_2\geq
0}\la_1^{n_1-n_2-2}\la_2^{n_2-1}q^{n_1+n_2-2}\eq$$
Manifestly, the projection of this function onto positive powers of $\la_1$ and $\la_2$ yields the desired
generating function once we set $\la_1=\la_2=1$. Let us introduce the
MacMahon projection symbol 
$\Omega$, defined by
$$\Ol\ \,\sum_{n=-\y}^\y c_n\la^n = \left. \sum_{n\geq0} c_n\la^n\right|_{\la=1}= \sum_{n\geq0} c_n\eq$$
For these projection manipulations, we use systematically 
 identities of the following type:
$$\eqalign{ {1\over (1-\la q) (1-\la^{-1} q)} &= {1\over (1-q^2)} \left({1\over
1-\la q} + {\la^{-1} q\over 1-\la^{-1}q}\right)\cr &=  {1\over (1-q^2)} \left({\la q\over
1-\la q} + {1\over 1-\la^{-1}q}\right)\cr}\eq$$
a proper choice of which being often an important simplifying factor (see for instance [\ref{L. B\'egin, C.
Cummins and P. Mathieu, J. Math. Phys. {\bf 41} (2000) 7640.}]).

Let us consider
$$\eqalign{
\Ola\Olaa \,G_1^{(2)} &= \Ola\Olaa\, {\la_1^{-2}\la_2^{-1}q^{-2}\over (1-\la_1 q)(1-\la_2 \la_1^{-1} q)}
=\Ola
\,{\la_1^{-3}q^{-1}\over (1-\la_1 q)(1- \la_1^{-1} q)}\cr &=\Ola\, {\la_1^{-3}q^{-1}}{1\over
(1-q^2)}\left({\la_1q\over (1-\la_1q)}+{1\over (1-
\la_1^{-1} q)}\right) \cr &=\Ola \,{\la_1^{-2}\over (1-q^2)(1-\la_1q)}  ={q^2\over (1-q)(1-q^2)}= {q^2\over
(q)_2}
\cr}\eq$$
(recall that after the $\la_i$ projection, we set $\la_i=1$). In the second line, we can drop the second term
because it only contains strictly negative powers of $\la_1$.  This is thus the generating function for the
$\Delta_2$-partitions into 2 parts. 

To count the number of  $\Delta_2$-partitions with $2j$ parts, we have to project the function
$$G_j^{(2)}= \la^{2j} q^{-2j^2}\prod_{i=1}^{2j}{\la_i^{-2}\over (1-\la_i\la_{i-1}^{-1}q)}\eq$$  
(with $\la_0=1$) onto non-negative powers of the various $\la_i$'s. If we ignore the powers of $\la_i$ in the
numerator, it amounts to neglect the $\Delta_2$ constraints as well as the strict positivity of the $n_i$'s.
The above projection would then lead directly to the generating function of the partitions of $n$ into at
most $2j$ parts, which reads $(q)_{2j}^{-1}$. The factors of $\la_i$ in the numerators are then easily  taken
into account: they simply generate the term $q^{2j^2}$ in the numerator, leading to the result:
$$\sum_s
p_{\Delta_2}^{[jk]}(s+2j^2) \, q^s = {q^{2j^2}\over (q)_{2j}}\eq$$
By summing over $j$, we recover the Lepowsky-Primc expression (\lepoo) for the Ising uncharged vacuum 
character, or equivalently, (\isingca) with $n\rw 2j$.

Consider now a somewhat less trivial example, the counting of the partitions  $p_{\Delta_3}^{[3]}(s+3)$. This
is associated to the following system of  Diophantine inequalities:
$$n_1\geq n_2\qquad n_2\geq n_3\geq 1\qquad n_1\geq n_3+2\eq$$
for $s+3=n_1+n_2+n_3$. 
The generating function for the solutions of this system is given by the projection of
$$G^{(3)}_1= {\mu^{-2}\la_3^{-1}q^{-3}\over (1-\mu\la_1 q)(1-\la_2 \la_1^{-1} q)(1-\la_3 \la_2^{-1}
\mu^{-1}q)}\eq$$
onto positive powers of the $\la_i$'s and of $\mu$. 
A simple analysis along the above lines leads to
$$\sum_n p_{\Delta_3}^{[3]}(s+3) \, q^s= {q^2(1+q-q^3)\over (q)_3} \eq$$
This can be compared to the restriction of (\lepoo) for $k=3$ in the 3-particle sector, which corresponds to
the terms $(m_1,m_2)$ with $m_1+2m_2=3$, namely $(3,0)$ and $(1,1)$; these  add up to
$$ {q^2\over (q)_1(q)_1}+ {q^6\over (q)_3} ={q^2(1+q-q^3)\over (q)_3}\eq$$
The coefficients of the $q$-expansion of this expression up to level 14 reproduce those listed in the
second row of (\decopa).

The higher-particle sectors can be treated by the same technique but the computations become quickly very
complicated.

\vskip0.3cm
\centerline{\bf Acknowledgment}

We thank professor J. Lepowsky for clarifying comments concerning his works and for pointing to us references [\LW,\DL]. This work was supported by NSERC.
\vskip0.3cm

%\vfill\eject
\centerline{\bf REFERENCES}
%\vskip 0.5cm
\immediate\closeout\refs \vskip 0.5cm
  \message{References}\input references
\vfill\eject

\end

----------------------------------------------
Coupures :

app C - finale

We have thus argued that the condition $(\psi_1)^k\sim \II$ amounts to eliminate one $k$-string at each level and that the coherence of this elimination process within longer strings  is ensured if the forbidden strings are chosen to have indices `as equal as possible'.  The linear independence of the states (\ferbas) subject to the restrictions (\restru) follows from the mere fact that there are no further sources of constrains. Admittedly, this last argument is heuristic.\foot{Note that in [\LP], the proof of linear independence is actually rather intricate.}

sur les v.s.

Indeed, the highest-weight modules constructed in terms of this  basis are highly reducible: there is an
infinite number of singular vectors. The two  ones are: 

These singular
vectors conditions imply the following constraints on the two states at the  extremes of the two strings
$\{\vph_{\ell}^{(r)}\}$ and $\{\vph_{\ell}^{(-r)}\}$: 
 $$\A_{-1}|\varphi^{(k-\ell)}_\ell\R=0 \, ,\qquad
\B_0|\varphi^{(-\ell)}_\ell\R=0\eqlabel\fert$$

..

Note at first that the field identifications reveal  the somewhat trivial nature of the parafermionic 
 singular vectors in the standard basis in that the singular vector conditions for
$|\vph_{\ell}\R$  are transformed in (\fert) into highest-weight conditions 
for $|\vph_{k-\ell}\R$.  However, the main point we want to make here is that

-------------

Consider first the $k=3$ case. Suppose that we act with the 3-string 
$$\A_{-{n_1}}\A_{-{n_2}}\A_{-{n_3}}= A_{-{\nu_1}}A_{-{\nu_2}}A_{-{\nu_3}}\qquad (\nu_i=n_i-f_i)\eqlabel\dede$$
on a state that does not contain $\A_{-n_3}$ excitations but which is otherwise arbitrary. 
Let $n_2=n_3$. According to the above rule, a third $\ZZ_3$ parafermionic excitation cannot be added
unless its energy is strictly greater that of the two $n_2$ modes, namely if
$$\nu_1 > \frac12(\nu_2+\nu_3)\quad \Rw n_1-f_1-\frac43 > \frac12(n_2-f_1-\frac23+n_2-f_1)\quad \Rw n_1>
n_2+1\eqlabel\troise$$
Similarly, consider the 3-string (\dede) with $n_1=n_2$ and look for the constraint on $n_3$: having
already two $\ZZ_3$ excitations of type $n_1$, we require the addition of a $\A_{-{n_3}}$ mode to be
allowed only if
$$\nu_3 < \frac12(\nu_1+\nu_2)\quad \Rw n_3-f_1 < \frac12(n_1-f_1-\frac43+n_1-f_1-\frac23)\quad \Rw n_3<
n_1-1\eqlabel\troise$$
These are exactly the same restrictions that follow from the effective $\ZZ_3$ exclusion principle
(\parexcl): the three 3-strings (\extri) have to be excluded. 

ancienne eq 4.12

:
$$\eqalign{
\chi_0^{(3)}& ={1\over \varphi(q)}[1-q+q^3-q^7-q^8+q^{14}-\cdots]\cr
&=\sum_{n=0}^\y  q^n[p(n)- p(n-1)+p(n-3)-p(n-7)-p(n-8)+p(n-14)-\cdots]\cr
&=[1+q^2+2q^3+3q^4+4q^5+7q^6+8q^7+ 12q^8+ 16q^9\cr &\quad + 22q^{10}+ 28q^{11}+ 39q^{12}+ 48q^{13}+ 65
q^{14}+\cdots]\cr}\eq$$

\subsec{The $k=4$ case}
As final simple example, consider the  uncharged vacuum module of the $\ZZ_4$ model that describes a free boson
on a rational torus of radius
$\sqrt{6}$. Up to level 9, the number of states in the 4- and 8-  particle states are given by
$$\matrix{
s&0&1&2&3&4&5&6&7&8&9\cr
p{[4]}&~&~&1&2&4&5&8&10&14&17\cr
p{[8]}&~&~&~&~&~&~&1&2&5&8\cr
}\eq$$
Here $p{[4j]}$ is the number of partitions of $s+4j^2$ into $4j$ terms that forbids the  four
4-strings (\exquad). For instance the 19 partitions associated to the $s=8$ level are:
$$\eqalign{
& (1119)\;\;(1128)\;\;(1137)\;\;(1146)\;\;(1155)\;\;(1227)\;\;(1236)\;\;(1245)\;\;(1335)\;\;\cr
&(1344)\;\;(2226)\;\;(2235)
\;\;(2244)\;\;(2334)\cr& (11133357)\;\;(11133366)\;\;(11133456)\;\;(11233455)\;\;(11133555)\cr}\eq$$
Again, these results can be compared with the corresponding vacuum Virasoro character multiplied by $q^{1/24}$
(see for instance [\DMS] p. 785) and denoted $\chi^{(4)}_0$:
$$\eqalign{
\chi_0^{(4)}= &{1\over 2\varphi(q)}\sum_{n\in\ZZ}(q^{3n^2}+(-1)^n q^{n^2})\cr
&=[1+q^2+2q^3+4q^4+5q^5+9q^6+12q^7+ 19q^8+ 25q^9+\cdots]\cr}\eq$$

section 5.2:

...................................
Let us apply this to the case of  the uncharged ($r=0$) character of $\vph_\ell$. Since we are
interested in counting 
$jk$-string partitions, we set $z^m=0$ if $m$ is not equal to $0$ mod $k$.  
For $m=jk$, we will fix $z^m$ in order to adjust the power of $q$ in our sum over $n$ to that of the level
$s$, namely, to transform
$$\sum_{n} p_{\Dkl}^{[m]}(s+kj^2+j\ell) \, q^nz^m \;\rw \;\sum_{s} p_{\Dkl}^{[jk]}(s+kj^2+j\ell) \,
q^s\eq$$
For this we need to set
$$ z^m=\left\{ \matrix{ q^{-m(m+\ell)/k}\qquad& {\rm
when}\quad m=jk \cr 0 &{\rm otherwise}\cr} \right.\eq$$ 
The identity (\ande) becomes then
$$\sum_{s,j=0}^\y p_{\Dkl}^{[jk]}(s+j^2k+j\ell) \, q^s =  \sum_{m_1,\cdots,m_{k-1}=0\atop 
N=0\;{\rm mod}\;k}^\y {
q^{N_1^2+\cdots+ N_{k-1}^2- N(N+\ell)/k+L_{k-\ell+1}}  \over (q)_{m_1}\cdots (q)_{m_{k-1}}}\eqlabel\lepo$$ 
On the rhs, the contribution to the $jk$-particle sector is selected by the
condition   $N = \sum im_i=jk$.

For the , 
we note  that  Hence here we put $z^m=0$ if $m$ is not equal to
$jk+r$ and adjust the power of $q$ as follows:
 $$ z^m=\left\{ \matrix{ q^{-(m-r)(m+r+\ell)/k -r}\qquad& {\rm
when}\quad m=jk +r \cr 0 &{\rm otherwise}\cr} \right.\eq$$ 
This  transforms (\ande) into
$$\eqalign{ &\sum_{s',j=0}^\y p_{\Dkl}^{[jk+r]}(s'+j^2k+j\ell+(2j+1)r) \, q^{s'} \cr
&\qquad \quad = 
\sum_{m_1,\cdots,m_{k-1}=0 \atop 
N=r\;{\rm mod}\;k}^\y { q^{N_1^2+\cdots+ N_{k-1}^2- (N-r)(N+r+\ell)/k-r+L_{k-\ell+1}}  \over
(q)_{m_1}\cdots (q)_{m_{k-1}} } \cr}\eqlabel\lepoo$$
Multiplied by $q^{h_\ell^{(r)}-c/24}$, this yields the general form of the $\ZZ_k$ charged characters in
a fermionic-sum representation. With this factor, this is precisely the
Lepowsky-Primc formula (theorem 9.4  of [\LP]).
......................................

apres C.7

That leads to an alternative expression for the constraint that exists at
every level, this times formulated in terms of a product of $k$ $\A$ operators.

appendix C

===============================

Let us detail the argument first for the case $k=3$. We start with (\commaa):
$$ \sum_{l=0}^{\infty} C^{(l)}_{1/3}  \A^{(2)}_{-n_1-n_{2}-l}
\A_{-n_3+l} \sim - \sum_{l=0}^{\infty} C^{(l)}_{1/3} \A_{-n_3-
l-1} \A^{(2)}_{-n_1-n_{2}+l+1}   
\eqlabel\commaab$$
We then develop the $ \A^{(2)}_m$ modes. Although there are many ways of separating the mode index
$m$ into two parts, we take one that selects the precise factor
$\A_{-n_1}\A_{-n_2}\A_{-n_3}$ as the lowest term of the double sum, by decomposing $
\A^{(2)}_{-n_1-n_{2}-l}$ on the lhs as follows:
$$\eqalign{ 
& c_{11} \sum_{l=0}^{\infty} C^{(l)}_{1/3}  \A^{(2)}_{-n_1-n_{2}-l}
\A_{-n_3+l} \cr & = \sum_{l,l'=0}^{\infty} C^{(l)}_{1/3} C^{(l')}_{-1/3} \left[  \A_{-n_1-l-l'}
\A_{-n_2+l'}+  \A_{-n_2-l'-1} \A_{-n_1-l+l'+1} \right]\A_{-n_3+l}\cr}\eq$$
The first double sum indeed contains the term $\A_{-n_1}\A_{-n_2}\A_{-n_3}$ as its $l=l'=0$ contribution. On
the rhs, we have
$$ \eqalign{ 
& c_{11}\sum_{l=0}^{\infty} C^{(l)}_{1/3} \A_{-n_3-
l-1} \A^{(2)}_{-n_1-n_{2}+l+1}\cr
=  &\sum_{l,l'=0}^{\infty} C^{(l)}_{1/3} C^{(l')}_{-1/3} \A_{-n_3-l-1}\left[  \A_{-n_1-l'+a}
\A_{-n_2+l+l'+1-a}+  \A_{-n_2+l-l'-a} \A_{-n_1+l'+1+a} \right]\cr}\eq$$
where $a$ is an arbitrary integer chosen to make the sum already as much ordered as possible. The choice
$a=0$ is convenient. Therefore,
$\A_{-n_1}\A_{-n_2}\A_{-n_3}$ can be expressed as a linear combination of terms of the following types
($l,l'\geq 0$):
$$ \eqalignD{ 
& (a)\quad  \A_{-n_1-l-l'} \A_{-n_2+l'} \A_{-n_3+l} \qquad &({\rm with}\; l+l'\geq 1)\cr
& (b)\quad  \A_{-n_2-l'-1} \A_{-n_1-l+l'+1}  \A_{-n_3+l} \qquad  &~\cr
& (c)\quad  \A_{-n_3-l-1} \A_{-n_1-l'} \A_{-n_2+l+l'+1}~&\cr
& (d)\quad  \A_{-n_3-l-1} \A_{-n_2-l'+l} \A_{-n_1+l'+1}~&\cr}\eq$$
on which no ordering has been done yet.

Let us concentrate on the first label. If the first
two terms are not already ordered, it means that the second label is larger (in absolute value)  than the first
one. Therefore, a lower bound condition on this first label cannot be violated by the reordering; it can only
be rendered weaker.  Comparing the first mode label of the four types of terms, we find two
conditions: $n_2+1\geq n_1$ and
$n_3+1\geq n_1$. The second inequality implies the first one.  We thus end up with the requirement 
$$n_3+1\geq n_1\eq$$
Suppose that $n_3=n_1+1$. The original triplet $(n_1,n_2,n_3)$ is thus either $(n_1,n_1,n_1-1)$ or
$(n_1,n_1-1,n_1-1)$. It is expressed in terms of a linear combination that contains $(n_1,n'_2,n'_3)$. But the
ordering condition and the sum constraint $n'_2+n'_3=n_2+n_3$ show that $n_2'=n_2$ and $n_3'=n_3$ is the only
solution.  Therefore, when $n_3+1\geq n_1$, the constraint on the first label ensures that
$\A_{-n_1}\A_{-n_2}\A_{-n_3}$ can be written in terms of terms that are distinct and also possibly
$\A_{-n_1}\A_{-n_2}\A_{-n_3}$ itself but no other terms in the set (\parexcl).

................

plus loin:

Take $k=3$ for instance and write down the states at the first few levels in the uncharged vacuum module:
$$\eqalign{
& s=0:\quad \A_{-1}\A_{-1}\A_{-1}\cr
& s=1:\quad \A_{-2}\A_{-1}\A_{-1}\cr
& s=2:\quad \A_{-3}\A_{-1}\A_{-1} \quad \A_{-2}\A_{-2}\A_{-1}\cr
& s=3:\quad \A_{-4}\A_{-1}\A_{-1} \quad \A_{-3}\A_{-2}\A_{-1}\quad \A_{-2}\A_{-2}\A_{-2}\cr
& s=4:\quad \A_{-5}\A_{-1}\A_{-1} \quad \A_{-4}\A_{-2}\A_{-1}\quad \A_{-3}\A_{-3}\A_{-1}\quad
\A_{-3}\A_{-2}\A_{-2}\cr}\eq$$ At each level we have to eliminate one state. We could select the one with
largest value of $n_1$.
===============================
------------------------------------------------------------
foot 1

\foot{Actually, the hint behind
the discovery of the explicit forms of the singular vectors is the following. Consider the application of a
string of $\A_{-1}$ operators on a highest state, say $(\A_{-1})^r| \varphi_\ell\R$.  When we keep
track of the fractional part, we can see that the dimension of the string increases as $r$
increases but at some point it will start to decrease;  when $r=k-\ell$ it has returned to
$h_{\varphi_\ell}$. Hence, in order to avoid sequences of operators that decrease the dimension, we
must forbid the state generated by the next application of $\A_{-1}$ on $(\A_{-1})^{k-\ell}|
\varphi_\ell\R$. This is thus the candidate singular vector. It can then be checked that this state is indeed a
parafermionic highest state, hence a genuine singular vector.  
A variant of this observation concerning the variation of the dimension of the operators along a string
provided us with a partial justification for the  formulation of our new basis of states (cf. the note in
section 5).} 

foot 2:

\foot{This difference condition was first obtained as a criterion 
 ensuring that the application of a $k$-string on a state does not decrease its  dimension.  A strong way of
ensuring this in full generality is to enforce the dimension of the right-most operator to be greater than
that of the left-most one.  Reinserting the fractional dimensions in the string acting on an arbitrary state
of charge $q$ (which can thus be a descendent)
$$\A_{-n_1}\cdots \A_{-n_k}| \phi_q\R= 
A_{-n_1+(1+2(k-1)+q)/k}\cdots A_{-n_k+(1+q)/k} |\phi_q\R$$
this crude criterion requires
$$n_1-{1+2(k-1)+q\over k} > n_k-{1+q\over k}\qquad \Rw \qquad n_1\geq n_k+2$$
The consideration of a similar string of $k$ operators inside any longer sequence  leads
to the above generalization.}

-----------------------------------------------  

Are there  analogous bosonic vs fermionic forms for the character expressions in the case of finite algebras?
Although a bit far-fetched, the following
$su(2)$ Lie-algebra example indicates that this is indeed so, at least if `fermionic representation' is
understood in the broader sense of `quasi-particle representation'.  For an irreducible representation with
finite Dynkin label
$\ell$ (i.e., $\ell= 2j$), which has dimension $\ell+1$, the Weyl character formula yields
$$\chi_{(\ell)} = { q^{-\ell}-q^{\ell+2} \over 1-q^{2} }= {q^{-\ell}(1-q^{2\ell+2)}\over (q^{2})_1 } $$
This is the bosonic form of the character, that captures the subtraction of one singular vector. On the other
hand, the Schwinger's oscillator model of angular momentum algebra, that is, the representation of the algebra
in terms of two independent harmonic oscillators (see e.g., [\ref{J.J. Sakurai, {\it Modern quantum
mechanics}, Addison Wesley (1994) sect. 3.8.}]),  yields
the quasi-particle representation:
 $$\chi_{(\ell)}= q^{-\ell} \sum_{m=0\atop m-\ell=0 \,{\rm mod} 2}^\ell q^{2m}= q^{-\ell} \sum_{m=0\atop
m-\ell=0 \,{\rm mod} 2}^\ell {(q^2)^m\over (q^2)_0}$$
This is obtained by projecting the product of the charged harmonic oscillator characters $(1-zq)^{-1}
\,(1-z^{-1}q)^{-1}$ onto the $z^{2\ell}$ sector. In this case, the quasi-particles have no fermionic
attribute.

\end

Let us look at special cases. For $k=2$, this should reduce to a free-fermionic basis. The $k=2$ version of
our basis is spanned by all states of the form
$$\A_{-n_1}\A_{-n_2}\cdots \A_{-n_m}|\varphi_\ell\R\qquad \quad {\rm with}\quad 
n_i\geq n_{i+1}+2\eqlabel\fermicas$$
In other words, in any ordered sequences of $\A$ modes, we exclude the following
two subsequences:
$$\A_{-n-1}\A_{-n} \qquad {\rm and} \quad \A_{-n}\A_{-n}\eq$$
If these act on a state of charge $q$, they read respectively
$$\eqalign{ A_{-n-1+(q+2)/2}A_{-n+q/2}&=  A_{-n+q/2}A_{-n+q/2}\cr
A_{-n+(q+2)/2}A_{-n+q/2}&= A_{-n+1+q/2}A_{-n+q/2}\cr}\eq$$
Hence, the forbidding of the first pair is simply a reflection of the usual Pauli exclusion
principle, or equivalently, the Grasmannian nature of the $\A$ modes.  On the other hand, the second
pair is omitted since it violates the ordering condition of the creation fermionic
operators.  In this simple case, that establishes the correctness of our criterion
(\restru).

For $k=3$, which pertains to three-state Potts model, the excluded states are all those
containing one of the following triplets
$$\A_{-n}\A_{-n}\A_{-n}\;\;,\qquad \A_{-n}\A_{-n+1}\A_{-n+1} \;,\qquad
\A_{-n}\A_{-n}\A_{-n+1}\eq$$

In general, the restriction (\restru) translates into the exclusion of all $k$-strings of
the form $(\A_{-n})^{k-i}(\A_{-n+1})^{i}$  for $i=0,\cdots , k-1$.
In other words, enforcing 
$$(\A_{-n})^{k-i}(\A_{-n+1})^{i}=0\qquad (i=0,\cdots , k-1)\eqlabel\parexcl$$
is an equivalent formulation of the parafermionic exclusion principle.

carac du vide:;
 we set
 $$ z^m=\left\{ \matrix{ q^{-m^2/k}\qquad& {\rm
when}\quad m=jk \cr 0 &{\rm otherwise}\cr} \right.\eq$$
Then, the sum over $m$ reduces to 
$$\sum_{m=0}^\y p_{\Dk}^{[m]}(n) \, q^nz^m = \sum_{j=0}^\y p_{\Dk}^{[jk]}(s+j^2k)\,  q^s\eq$$
With this choice for $z$, in the rhs of (\ande) we get
$$ z^{N}=\left\{ \matrix{ q^{-(N^2/k} \qquad& {\rm
when}\quad N=jk \cr 0 &{\rm otherwise}\cr} \right.\eq$$
The identity (\ande) becomes then
$$\sum_{s,j=0}^\y p_{\Dk}^{[jk]}(s+j^2k) \, q^s = \left. \sum_{m_1,\cdots,m_{k-1}=0}^\y { q^{N_1^2+\cdots+
N_{k-1}^2- N^2/k}  \over (q)_{m_1}\cdots (q)_{m_{k-1}} }\right|_{N=0\;{\rm
mod}\;k}\eqlabel\lepo$$  On the rhs, the contribution to the $jk$-particle sector is selected by the
condition   $N=\sum N_i = \sum im_i=jk$. Up to an overall factor $q^{-c/24}$, this is precisely the
Lepowsky-Primc formula for the vacuum character (theorem 9.4  of [\ref{J. Lepowsky and M. Primc,  Contemporary
Mathematics {\bf 46} AMS, Providence, 1985.}\refname\LP]).  
}

Rocha-Caridi...
$$
\chi_{(r,s)}(q)={q^{-{c/24}} \over \varphi(q)} 
\bigg[q^{\displaystyle h_{r,s}} +\sum_{k=1}^{\infty} 
(-1)^k \big\{&q^{\displaystyle
h_{\SCR r+kp',(-1)^ks+[1-(-1)^k]p/2}} \cr
&+q^{\displaystyle
h_{\SCR r,kp+(-1)^k s+[1-(-1)^k]p/2}} \big\} \bigg]
\cr}\nom{irchar}$$

$$K_{r,s}^{(p,p')}(q)= { q^{-{1/24}} \over \varphi(q)} 
\sum_{n \in \ZZ} q^{(2pp'n+ pr- p's)^2/4pp'}\nom{Kfunct}$$
as
$$ \chi_{(r,s)}(q)= { q^{-{1/24}} \over (q)_\y} \sum_{n \in \ZZ} [q^{(2pp'n+ pr- p's)^2/4pp'}- \sum_{n \in
\ZZ} q^{(2pp'n+ pr+ p's)^2/4pp'}]
\eqlabel\rocha$$

% $$\matrix{
% &s=2: \; &(13)\;&\;(35)\;&\;\;&\;\;&\;\;&\;\;&\;\;&\;\cr
% &s=3:  \;&(14) \;&\;(35)\;&\;\;&\;\;&\;\;&\;\;&\;\;&\;\cr
% &s=4:  \; &(15)\;&\;(24)\;&\;(35)\;&\;\;&\;\;&\;\;&\;\;&\;\cr 
% &s=5:  \; &(16)\;&\;(25)\;&\;(35)\;&\;\;&\;\;&\;\;&\;\;&\;\cr
% &s=6:  \; &(17)\;&\;(26)\;&\;(35)\;&\;\;&\;\;&\;\;&\;\;&\;\cr
% &s=7:  \; &(18)\;&\;(27)\;&\;(36)\;&\;\;&\;\;&\;\;&\;\;&\;\cr
% &s=8: \; &(19)\;&\;(28)\;&\;(37)\;&\;(46)\;&\; (1357) \;&\;\;&\;\;&\;\cr
% &s=9: \; &(1,10)\;&\;(29)\;&\;(38)\;&\;(47)\;&\; (1358)\;&\;\;&\;\cr
% &s=10: \;&(1,11)\;&\;(2,10)\;&\;(39)\;&\;(48)\;&\;(57)\;&\; (1359)\;&\;(1368)\cr
% }\eq$$